\documentclass[12pt]{article}

\usepackage{cite}
\usepackage{ctex}
\usepackage{times}
\usepackage{mathrsfs}
\usepackage{graphicx}
\usepackage{bm}
\usepackage{float}
\usepackage{physics}
\usepackage{siunitx}
\usepackage{xcolor}
\usepackage{subfigure}
\usepackage{appendix}
\usepackage[colorlinks,linkcolor=blue,anchorcolor=blue,urlcolor=blue,citecolor=blue]{hyperref}
\usepackage{multirow}
\usepackage{epstopdf}
\usepackage{soul}
\usepackage{bbm}
\usepackage{booktabs}
\usepackage{colortbl}

\newenvironment{sciabstract}{%
	\begin{quote} \bf}
	{\end{quote}}

\newcounter{lastnote}

\title{Experimental Realization of a Nonreciprocal Quantum Battery}

\author
{Jie Zhang$^{1,2\dagger}$, Xin-Sheng Chen$^{1,2\dagger}$, Ruo Kun Cai$^{1,2}$, Gang Xia$^{1,2}$, \\
	Jia-Yang Song$^{1,2}$, Fu-Wen Zhao$^{1,2}$, Ting Chen$^{1,2}$, Wei Wu$^{1,2}$, Yi Xie$^{1,2\ast}$,\\
	Zhi Jiao Deng$^{1,2\ast}$, Chun-Wang Wu$^{1,2\ast}$, Ping-Xing Chen$^{1,2,3}$\\
	\normalsize{$^{1}$College of Science, National University of Defense Technology,}\\
	\normalsize{Changsha 410073, Hunan, China} \\
	\normalsize{$^{2}$Hunan Key Laboratory of Quantum Information Mechanism and Technology,}\\
	\normalsize{NUDT, Changsha 410073, Hunan, China}\\
	\normalsize{$^{3}$Hefei National Laboratory, Hefei 230088, Anhui, China}\\
	\\
	\normalsize{$^\dagger$These authors contributed equally to this work.}\\
	\normalsize{$^\ast$Corresponding authors:}\\
	\normalsize{Email: xieyi2015@nudt.edu.cn (Y.X.); dengzhijiao926@hotmail.com (Z.J.D.);}\\
	\normalsize{cwwu@nudt.edu.cn (C.W.)}
}

\date{}

\begin{document}
	
	\baselineskip24pt
	
	\maketitle
	

\begin{sciabstract}
	
	Quantum batteries promise faster and more efficient energy storage and extraction by harnessing quantum mechanics, serving as essential building blocks for future quantum devices. A key practical requirement—unidirectional energy flow during charging—is straightforward in classical batteries via potential difference, yet remains a major challenge in the quantum regime due to the inherent reciprocity of coherent coupling. Here, we realize the first backflow-free, nonreciprocal quantum battery in the few-quantum regime, implemented in a highly controlled trapped-ion system via a qubit-reservoir engineering technique. By subtly tailoring quantum interference between coherent and dissipative couplings of two laser-cooled motional modes, we achieve strictly unidirectional energy flow from the charger to the battery. We observe maximum charging power at an exceptional point and demonstrate a complete charge–discharge cycle with unitary work extraction. This work establishes the foundation for directional energy transfer and storage in micro- and nanoscale quantum devices, with broad implications for nonreciprocal energy harvesting, precision metrology, information processing, and quantum thermodynamics.
\end{sciabstract}

Quantum batteries \cite{zhongshu2024, zhongshu2026, Virgili2025} leverage coherence and entanglement for energy storage in quantum systems, offering potential advantages over their classical counterparts and positioning themselves as key components for future quantum technologies \cite{Quach22, Quach26, computer2026, Alicki_Fannes2013, Binder2015, Campaioli17, Ferraro18, Rossini20, Farina19, Gherardini20, Song24, Tibben25, Lu25, Song25}. While quantum coherence underlies the advantageous features of quantum batteries, purely coherent charging suffers from transient energy backflow due to the time-reversal symmetry of coherent coupling \cite{Andolina18, Joshi22, Crescente24, Andolina25}. In contrast, classical electrochemical batteries readily achieve unidirectional energy flow by maintaining a higher voltage at the charger. In the quantum realm, backflow can in principle be eliminated by switching off the coupling at the moment of full charge, but this is infeasible in many practical scenarios as it requires complete prior knowledge of both the initial state and the evolution time. When this stringent requirement is not satisfied (e.g., for a used battery with unknown residual energy, or a battery to be charged with a mixed initial state due to noise), deterministic charging becomes impossible and the optimal switching instant cannot be determined. Furthermore, it has been shown that environmental noise severely degrades the capacity of purely coherently charged batteries \cite{Ahmadi24}. These challenges motivate new designs that enable stable charging without switching off the coupling while remaining insensitive to noise \cite{Ahmadi24, Barra19, Santos19}.


Recently, nonreciprocity has been introduced into quantum batteries to address these challenges \cite{Ahmadi24}. This approach exploits quantum interference between coherent and dissipative couplings, with the latter mediated by a shared reservoir, to break time-reversal symmetry \cite{Metelmann15}. It thereby enables unidirectional energy flow from the charger to the battery without precise timing control or state tracking, while exhibiting robustness against noise  \cite{Ahmadi24}. It is worth noting that reservoir-engineered nonreciprocity has been demonstrated in optomechanical systems \cite{Fang17, Bernier17, Peterson17, Xu19}, atomic ensembles \cite{Lu21}, and cavity magnonics \cite{Wang19}, but exclusively for signal or energy transmission operating deep in the classical regime. Realizing a nonreciprocal quantum battery, where quantum coherence and fluctuations dominate, requires pushing this technique into the few-excitation regime, an experimentally unexplored frontier.

Here, we experimentally realize a nonreciprocal quantum battery in a single trapped calcium ion with an average phonon number slightly above 3. Unlike previous schemes employing harmonic-oscillator reservoirs \cite{Fang17, Bernier17, Peterson17, Xu19, Lu21, Wang19}, a qubit reservoir is implemented here using Jaynes-Cummings (JC)-type quantum control. The experiment verifies phase-controlled directional energy transfer between the charger and the battery and exploits it for nonreciprocal charging without energy backflow. The charging power reaches its maximum at an exceptional point (EP), and unitary work extraction from the steady state is demonstrated. This work reports the first experimental realization of nonreciprocal charging in the quantum regime, validating reservoir-engineered directional energy storage.

\begin{figure*}[htbp]
	\centering
	\includegraphics*[scale=0.75]{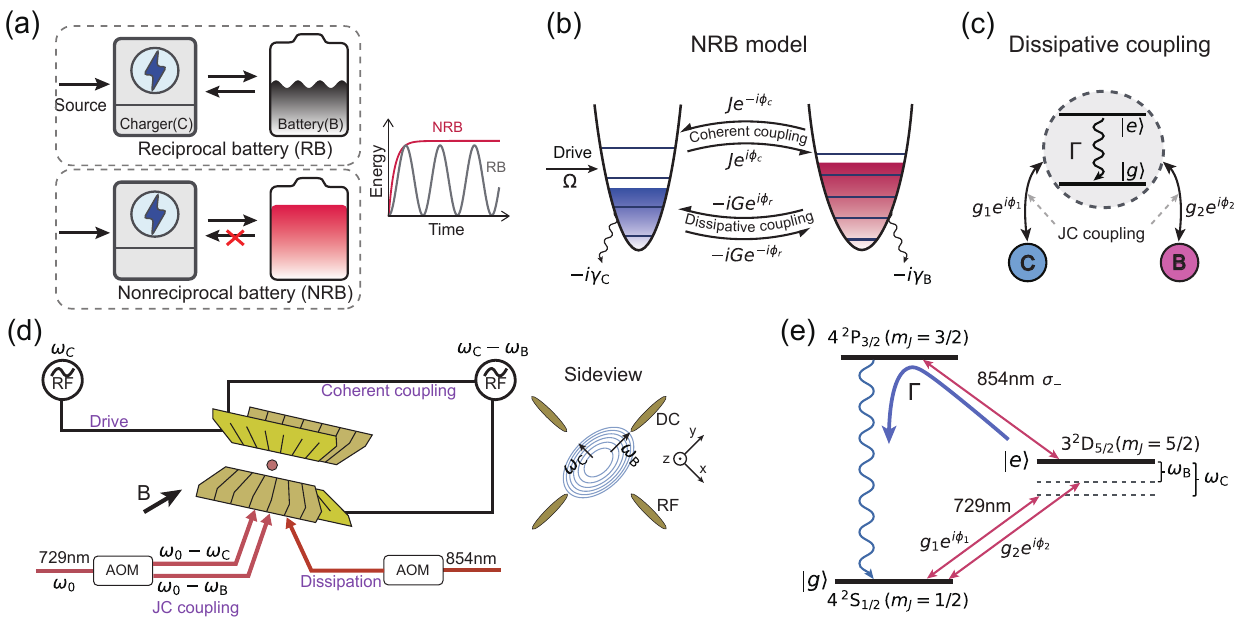}
	\caption{\textbf{Nonreciprocal quantum battery: from concept to trapped-ion realization.} \textbf{(a)} Comparison of charging in an RB and an NRB. In the RB, bidirectional energy flow causes the battery energy to oscillate, whereas in the NRB, unidirectional flow causes it to increase monotonically to a steady value. \textbf{(b)} Theoretical model of the NRB. The driven charger mode interacts with the battery mode via coherent and dissipative couplings, the latter introducing decay rates $\gamma_C$ and $\gamma_B$. \textbf{(c)} Schematic implementation of the dissipative coupling. Both modes are coupled to a common two-level qubit bath via JC couplings, with the qubit excited state decaying at a rate $\Gamma$. \textbf{(d)} Experimental setup and control scheme. A single $^{40}\mathrm{Ca}^{+}$ ion is trapped in a segmented-blade Paul trap. RF drives and optical lasers are applied to realize motional driving, coherent coupling, and dissipative coupling. Inset: sideview of the radial trapping pseudopotential. \textbf{(e)} Energy levels and optical transitions. A bichromatic 729 nm laser drives the red-sideband transitions for the JC coupling, while an 854 nm laser induces an effective decay of $|e\rangle$.		
	}
	\label{figure1}
\end{figure*}

The concept of nonreciprocal charging originates from the nonreciprocal quantum battery proposed by Ahmadi et al. \cite{Ahmadi24}. In contrast to a reciprocal battery (RB) exhibiting bidirectional energy flow, a nonreciprocal battery (NRB) permits only unidirectional flow from charger to battery (Fig.~\ref{figure1}a). In their scheme, two harmonic oscillators act as the charger and the battery, with the former externally driven and their interaction mediated by both coherent and dissipative couplings (Fig.~\ref{figure1}b). The mode amplitudes obey the following equations of motion:
\begin{equation}\label{coupled}
	\begin{split}
		\frac{d\langle \hat{a}_1 \rangle}{dt} &= -i \left( \frac{J}{2} e^{-i\phi_c} - i G e^{i\phi_r} \right) \langle \hat{a}_2 \rangle - \gamma_C \langle \hat{a}_1 \rangle - i \Omega e^{i\phi_d}, \\
		\frac{d\langle \hat{a}_2 \rangle}{dt} &= -i \left( \frac{J}{2} e^{i\phi_c} - i G e^{-i\phi_r} \right) \langle \hat{a}_1 \rangle - \gamma_B \langle \hat{a}_2 \rangle.
	\end{split}
\end{equation}
Here, $\langle \hat{a}_1 \rangle$ and $\langle \hat{a}_2 \rangle$ are the mean amplitudes of the charger and battery modes, respectively. $J$ ($G$) and $\phi_c$ ($\phi_r$) denote the strength and phase of the coherent (dissipative) coupling. The charger and battery decay at rates $\gamma_C$ and $\gamma_B$, respectively, with an external drive of amplitude $\Omega$ and phase $\phi_d$ applied to the charger. The required unidirectionality arises from the destructive interference between the coherent and dissipative couplings, which cancels the backaction from the battery to the charger for $J/2 = G$ and $\phi_c + \phi_r = -\pi/2$.

A crucial step is the construction of the dissipative coupling (Fig.~\ref{figure1}c). Conventionally, it is realized by coupling both modes to a common harmonic-oscillator bath. Here, we replace this bath with a two-level qubit bath, leveraging JC coupling control techniques in a trapped-ion platform. The coupling between the qubit and mode $j$ is given by $\frac{g_j}{2}(e^{-i\phi_j}\hat{\sigma}_+ \hat{a}_j + e^{i\phi_j}\hat{\sigma}_- \hat{a}_j^\dagger)$ ($j=1,2$), where $\hat{\sigma}_\pm$ are the qubit raising and lowering operators. This scheme requires only the internal levels and motional modes of a single ion. For an excited-state decay rate $\Gamma \gg g_1, g_2$, adiabatic elimination of the internal states yields an effective dissipative coupling between the charger and battery modes, with $G = \frac{g_1 g_2}{2\Gamma}$, $\gamma_C = \frac{g_1^2}{2\Gamma}$, $\gamma_B = \frac{g_2^2}{2\Gamma}$, and $\phi_r = \phi_1 - \phi_2$ (see Supplementary Materials).

Our experiment uses a single $^{40}\mathrm{Ca}^{+}$ ion trapped in a segmented-blade Paul trap (Fig.~\ref{figure1}d). RF and DC signals on the light- and dark-yellow electrodes provide radial and axial confinement, respectively. The ion's motional modes are well approximated as harmonic oscillators, and the $x$ and $y$ radial modes serve as the charger and battery modes. The two modes have trap frequencies of $\omega_C/2\pi \approx 2.71\,\mathrm{MHz}$ and $\omega_B/2\pi \approx 2.38\,\mathrm{MHz}$, and motional coherence times exceeding $6\,\mathrm{ms}$ and $11\,\mathrm{ms}$, respectively. A static magnetic field of about $7.35\,\mathrm{G}$, oriented at roughly $45^\circ$ to the trap axis, defines the quantization axis and induces a Zeeman splitting of the relevant sublevels. The qubit is encoded in $|g\rangle \equiv |4s\ ^{2}\text{S}_{1/2}, m_J=1/2\rangle$ and $|e\rangle \equiv |3d\ ^2D_{5/2}, m_J=5/2\rangle$, and the $|g\rangle \leftrightarrow |e\rangle$ quadrupole transition is driven by a $729\,\mathrm{nm}$ laser with a linewidth near $10\,\mathrm{Hz}$.

To engineer the effective Hamiltonian, we combine electrical and optical control of the trapped ion (Figs.~\ref{figure1}d and \ref{figure1}e). Electrically, a resonant RF signal at $\omega_C$ applied to one trap blade drives the charger mode via the term $\Omega ( e^{i\phi_d} \hat{a}_1^\dagger + e^{-i\phi_d} \hat{a}_1 )$, and two identical RF signals at $\omega_C - \omega_B$ on the diagonal blades induce a coherent coupling $\frac{J}{2} (e^{-i\phi_c}\hat{a}_1^\dagger\hat{a}_2 + e^{i\phi_c}\hat{a}_1\hat{a}_2^\dagger)$. Optically, a $729\,\mathrm{nm}$ laser resonant with the qubit transition at frequency $\omega_0$ is modulated by an acousto-optic modulator (AOM) to generate two red-sideband frequency components at $\omega_0 - \omega_C$ and $\omega_0 - \omega_B$, which couple the internal states $|g\rangle$ and $|e\rangle$ to the charger and battery modes with coupling strengths $g_1 e^{i\phi_1}$ and $g_2 e^{i\phi_2}$, respectively, realizing the JC coupling. Meanwhile, a circularly polarized $854\,\mathrm{nm}$ laser couples $|e\rangle$ to the short-lived $|4p\ ^2P_{3/2}, m_J=3/2\rangle$ state, followed by spontaneous emission back to $|g\rangle$ to induce an effective decay whose rate is controlled by adjusting the $854\,\mathrm{nm}$ laser intensity. The JC coupling and this effective decay together yield the dissipative coupling.

The model requires precise frequency matching among the lasers, electrical drives, atomic levels, and motional modes, whereas trap-frequency drift and Stark shifts introduce detunings. Simulations show that the drive detuning can be tolerated up to $20\,\mathrm{Hz}$, while the coherent coupling and red-sideband detunings must be kept below $5\,\mathrm{Hz}$. We compensate for the Stark shift of the $|e\rangle$ level induced by the intense $854\,\mathrm{nm}$ laser by tuning the 729 nm laser frequency to optimize the sideband cooling, which relies on the red-sideband JC coupling and the laser-induced dissipation. Trap-frequency drift is actively suppressed, as detailed in the Supplementary Materials.  To further suppress residual drift, we interleave calibration and data acquisition: the motional-mode resonance frequencies are calibrated by sweeping the frequency of weak probe signals and monitoring the mode excitation response,  and then the drive frequencies are updated for subsequent experimental runs. This dynamic calibration keeps the drive detuning below $20\,\mathrm{Hz}$ and the coherent coupling and red-sideband detunings below $5\,\mathrm{Hz}$, ensuring the destructive interference required for nonreciprocal charging.

\begin{figure*}[htbp]
	\centering
	\includegraphics*[scale=0.75]{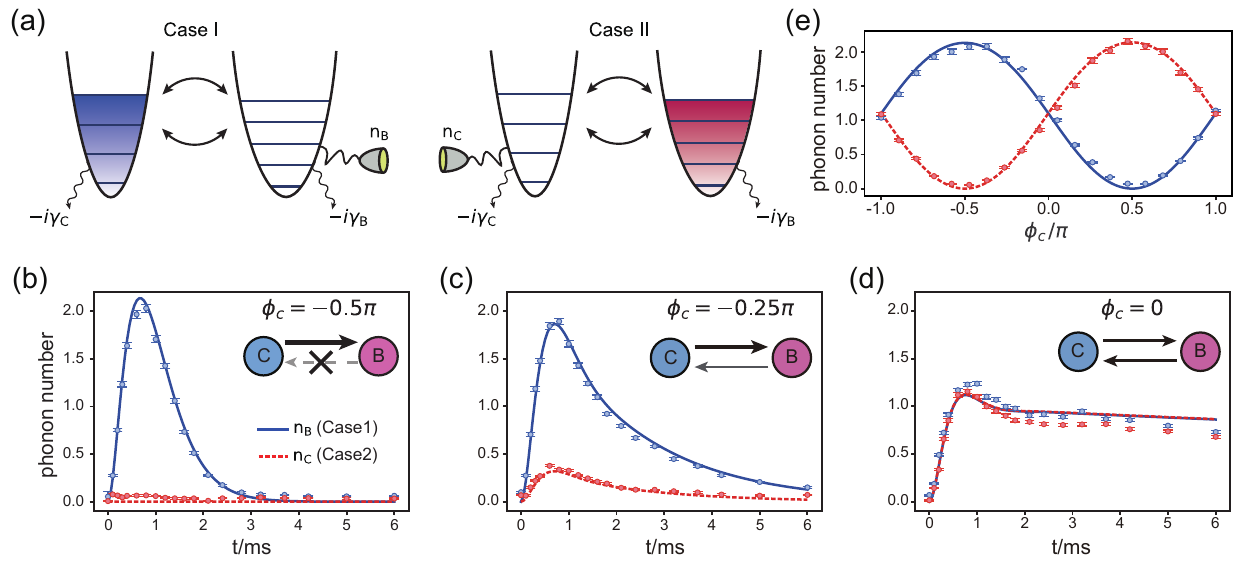}
	\caption{\textbf{Phase-controlled directionality of energy transfer.} \textbf{(a)} Schematics of the two configurations. Case I: energy initially stored in the charger, battery probed as the target mode. Case II: roles reversed. \textbf{(b--e)} Target-mode mean phonon number, with (b--d) showing its time evolution for three representative values of $\phi_c$ and (e) showing its phase dependence at $t = 0.65$ ms. Blue and red points represent experimental data for Cases I and II, respectively, and solid and dashed lines are theoretical simulations. Error bars represent the standard deviation propagated from projection-measurement noise. The parameters are $g_1/2\pi = 5$ kHz, $g_2/2\pi = 6$ kHz, $\Gamma/2\pi = 64$ kHz, and $J/2\pi = 0.469$ kHz with the initially energy-storing mode prepared in a coherent state $|\alpha = 2.0\rangle$.
	}
	\label{figure2}
\end{figure*}

As a prerequisite for nonreciprocal charging, we first verify the core energy transfer mechanism in the absence of external driving. Achieving zero backflow requires conditions on both the strengths of the coherent and dissipative couplings and their relative phases. In our experiment, we set $J/2 = G$ and $\phi_r=0$, and solely tune the phase $\phi_c$ of the RF drive used for coherent coupling. We initialize one mode with a coherent state of mean phonon number $\bar{n} \approx 4$ while leaving the other in vacuum, and monitor the energy evolution of the latter, hereafter referred to as the target mode. Two configurations are compared (Fig.~\ref{figure2}a): energy is initially stored in the charger for Case I and in the battery for Case II. Exchanging the roles of the two modes directly probes whether the system prefers a direction of energy transfer. 

Experimentally, we first calibrate the system parameters to achieve $J/2 = G$. Both motional modes are laser cooled to near their ground states, and one mode is then electrically driven to a coherent state with $\bar{n} \approx 4$. Subsequently, an RF drive at $\omega_C - \omega_B$, an $854\,\mathrm{nm}$ and a bichromatic $729\,\mathrm{nm}$ laser beams are applied simultaneously. After an evolution time, the mean phonon number of the target mode is extracted from the phonon number distribution obtained by fitting blue-sideband oscillations. The phase origin is calibrated by identifying the $\phi_c$ that minimizes the backflow and matching it to the theoretically predicted phase.

The measured time evolution of the target-mode mean phonon number reveals a clear dependence on $\phi_c$ (Fig.~\ref{figure2}b--d). Because the dissipative coupling damps both modes, the mean phonon number typically builds up from zero to a maximum and then decays back to zero. As $\phi_c$ increases from $-0.5\pi$ to $0$, the unidirectionality weakens, and the energy transfer becomes symmetric at $\phi_c = 0$. Fig.~\ref{figure2}e quantifies this asymmetry at $t = 0.65$ ms, near the peak of the time evolution. At $\phi_c = -0.5\pi$, the target mode is maximally excited in Case I, whereas it is essentially unexcited in Case II at this time. This absence of excitation is consistent with the zero backflow observed in the full time evolution (Fig.~\ref{figure2}b). At $\phi_c = 0.5\pi$, the roles are reversed, indicating unidirectional flow from battery to charger throughout the transfer. The curves intersect at $\phi_c = 0$ and $\pm\pi$, across which the dominant transfer direction flips.

\begin{figure}[htb]
	\centering
	\includegraphics[scale=0.75]{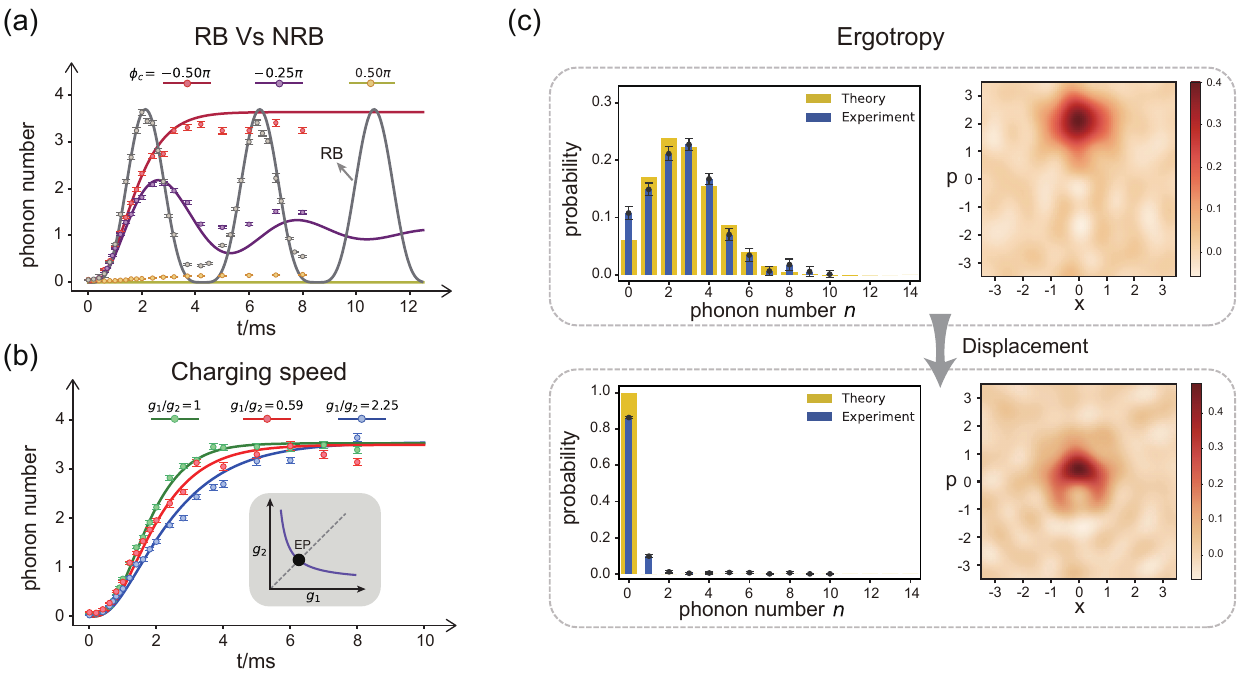}
	\caption{\textbf{Nonreciprocal charging dynamics and work extraction.} \textbf{(a)} Time evolution of the battery mean phonon number for different phases $\phi_c$, with the RB shown for comparison. Solid lines and discrete points represent theoretical simulations and experimental data, respectively. The optimal phase ($\phi_c = -0.50\pi$) realizes the NRB, whose steady-state mean phonon number is slightly below the RB peak, attributed to the imperfect adiabatic approximation. The parameters are $g_1/2\pi = 5$ kHz, $g_2/2\pi = 6$ kHz, and $\Omega/2\pi = 0.225$ kHz. \textbf{(b)} Charging dynamics measured for different ratios of $g_1/g_2$ while keeping the product $g_1 g_2$ fixed. The data show the time evolution of the battery mean phonon number for three representative ratios. Inset: parameter space illustrating the EP at $g_1 = g_2$. For this EP case, $g_1/2\pi = g_2/2\pi = 5.46$ kHz and $\Omega/2\pi = 0.22$ kHz. \textbf{(c)} Measured phonon number distributions and Wigner functions of the battery mode before (top) and after (bottom) the displacement operation. The pre-displacement state is measured after 4 ms of charging dynamics from the ground state, by which time the steady state has been reached. The parameters are $g_1/2\pi = g_2/2\pi = 5.9$ kHz and $\Omega/2\pi = 0.23$ kHz. In all panels, error bars represent the standard deviation propagated from projection-measurement noise, and $\Gamma/2\pi = 64$ kHz, $J = g_1 g_2 / \Gamma$.}
	\label{figure3}
\end{figure}

Building on the nonreciprocal energy transfer mechanism demonstrated in Fig.~\ref{figure2}, we experimentally investigate nonreciprocal charging by applying an additional resonant RF drive to the charger and initializing both modes in their ground states. At the optimal phase ($\phi_c = -0.5\pi$), the battery mean phonon number in Fig.~\ref{figure3}a rises monotonically and stabilizes near the maximum value reached during the coherent oscillation without dissipative coupling, eliminating the need for precise timing control. This stabilization originates from the cascaded dynamics captured by the lower-triangular coefficient matrix 
\begin{equation}\label{dynamicmatrix}
	\mathbf{M} = \begin{pmatrix} -\gamma_C & 0 \\ -iJ e^{i\phi_c} & -\gamma_B \end{pmatrix},
\end{equation}
which enforces unidirectional driving that brings the battery to a steady state. In the ideal adiabatic limit, this steady state is a coherent state whose amplitude $\langle \hat{a}_2(\infty) \rangle = -4\Omega e^{i(\phi_c+\phi_d)}/J$ depends solely on \(\Omega\) and \(J\), and coincides with the peak amplitude of the coherent oscillation (see Supplementary Materials). At $\phi_c = -0.25\pi$, the data show transient oscillations before settling into a steady state with a lower mean phonon number, and at $\phi_c = 0.5\pi$, the unidirectionality reverses, preventing energy from entering the battery. The deviation of the experimental data from the theoretical curves at later evolution times originates from accumulated detuning caused by frequency drift.

The unidirectional charging speed exhibits a strong dependence on the balance between the decay rates. By tuning the ratio $g_1/g_2$ while keeping their product constant, we keep the steady-state energy fixed and vary the decay rates $\gamma_C$ and $\gamma_B$. As shown in Fig.~\ref{figure3}b, the charging dynamics reveal that the steady state is reached fastest at $g_1 = g_2$, corresponding to $\gamma_C = \gamma_B$. We identify this point as an EP with critical damping, where the two eigenmodes coalesce. This observation can be explained by the cascaded process: the relaxation time is limited by the slower of the two decay rates, and the shared reservoir together with the nonreciprocal condition fixes their product to $J^2/4$, so the fastest charging occurs at the EP, providing a direct experimental route to optimal nonreciprocal charging power.

To verify the coherent nature of the steady state and quantify the extractable work, we apply a unitary displacement operation to shift the battery mode by an amount $-\langle \hat{a}_2(\infty) \rangle$, and measure the phonon number distribution and Wigner function before and after the displacement (Fig.~\ref{figure3}c). The Wigner function is reconstructed from the measured characteristic function via a two-dimensional Fourier transform (see Supplementary Materials). Before displacement, the Wigner function exhibits the characteristic shape of a displaced vacuum, corresponding to a coherent state. After displacement, the state is close to vacuum, with an $n=0$ probability of about 0.86. The measured mean phonon numbers before and after the displacement are $\bar{n}_{\mathrm{steady}} \approx 2.74$ and $\bar{n}_{\mathrm{after}} \approx 0.27$, with the residual excitation mainly due to experimental imperfections. This confirms that the stored energy is predominantly coherent and can be extracted by a unitary displacement operation, thereby completing the discharge step.



By engineering dissipative coupling through a qubit reservoir in a single trapped ion, we achieved unidirectional energy flow from the charger to the battery, eliminating the backflow limiting conventional coherent charging. At the optimal relative phase between the coherent and dissipative couplings, the unidirectional flow drives the battery to a steady state matching the peak of the coherent oscillation. We further identified an exceptional-point condition maximizing charging power and verified unitary extractability of the stored energy. These results offer a pathway toward practical quantum batteries and chiral energy networks in nano- or microscale quantum devices, as well as the integration of nonreciprocal energy routing with other quantum components. Moreover, the extraordinary sensitivity of the battery response to experimental parameters, particularly the driving-field frequencies, implies great potential for quantum metrology \cite{Lau18, Dechant25}. In a broader context, our platform provides an ideal testbed for studying information theory and quantum thermodynamics related to quantum nonreciprocal phenomena.

\providecommand{\noopsort}[1]{}\providecommand{\singleletter}[1]{#1}%

\section*{Acknowledgments}
\noindent{We thank Chuan-Feng Li and Jun-Hong An for carefully reading the manuscript and providing valuable suggestions. This work was supported by the National Natural Science Foundation of China (Grants No. 12174447, No. 12174448, No. 12204543, No. 11904402 and No. 11574398), Quantum Science and Technology-National Science and Technology Major Project (QNMP) (Grant No. 2021ZD0301605), the Science and Technology Innovation Program of Hunan under Grant No. 2022RC1194, the Natural Science Foundation of Hunan Province under Grant No. 2023JJ10052, and the Innovation Research Foundation of NUDT.} 

\noindent{\textbf{Author contributions:}} {P.C. supervised the entire project. C.W., J.Z., Z.J.D., and P.C. conceived the research. Z.J.D. and C.W. performed the theoretical analysis. C.W. and J.Z. designed the experiment. J.Z., X.C., and Y.X. carried out the experiments. J.Z., X.C., and R.K.C. analyzed the data. R.K.C. prepared the figures. G.X., J.S., F.Z., T.C., and W.W. contributed to building the experimental platform. C.W., Z.J.D., and J.Z. wrote the manuscript. All authors contributed to discussing the results and reviewing the manuscript.}

\noindent{\textbf{Competing interests:}} The authors declare no competing financial interests.

\noindent{\textbf{Data and materials availability:}} All data needed to evaluate the conclusions in the paper are present in the paper and/or the Supplementary Materials.

\clearpage

	\clearpage
	
	\section*{Supplementary Materials for \\
		\bfseries Experimental Realization of a Nonreciprocal Quantum Battery}
	
	\setcounter{section}{0}
	\renewcommand{\thesection}{S\arabic{section}}
	
	\setcounter{equation}{0}
	\renewcommand{\theequation}{S\arabic{equation}}
	
	\setcounter{figure}{0}
	\renewcommand{\thefigure}{S\arabic{figure}}
	
	\setcounter{table}{0}
	\renewcommand{\thetable}{S\arabic{table}}

%
%

%
%
%
%
\textbf{This PDF file includes:} Supplementary Text; Figs. S1 to S12.

In the Supplementary Materials, we present a detailed derivation of the effective master equation for the nonreciprocal quantum battery model, along with a mean-field analysis and numerical simulations of the dynamics, as well as the influence of detuning errors. Additionally, we provide a comprehensive description of the trapped-ion experimental apparatus, the calibration procedures for relevant parameters, and the reconstruction techniques for both the Fock-state probabilities and the phase-space Wigner function of the motional state.

\section{Derivation of the effective master equation}

We consider a single trapped ion whose internal state is modeled as a two-level system with ground state \(|g\rangle\) and excited state \(|e\rangle\), and whose two radial motional modes act as the charger and the battery, with annihilation operators \(\hat{a}_1\) and \(\hat{a}_2\), respectively. In the interaction picture, the total Hamiltonian is
\begin{equation}
	\begin{aligned}
		\hat{H}_{\mathrm{full}}&=\hat{H}_{\mathrm{drive}}+\hat{H}_{gC}+\hat{H}_{gB}+\hat{H}_{\mathrm{coup}},\\
		\hat{H}_{\mathrm{drive}}&=\Omega\left(e^{i\phi_d}\hat{a}_1^\dagger+e^{-i\phi_d}\hat{a}_1\right),\\
		\hat{H}_{gC}&=\frac{g_1}{2}\left(e^{-i\phi_1}\hat{\sigma}_+\hat{a}_1+e^{i\phi_1}\hat{\sigma}_-\hat{a}_1^\dagger\right),\\
		\hat{H}_{gB}&=\frac{g_2}{2}\left(e^{-i\phi_2}\hat{\sigma}_+\hat{a}_2+e^{i\phi_2}\hat{\sigma}_-\hat{a}_2^\dagger\right),\\
		\hat{H}_{\mathrm{coup}}&=\frac{J}{2}\left(e^{-i\phi_c}\hat{a}_1^\dagger\hat{a}_2+e^{i\phi_c}\hat{a}_1\hat{a}_2^\dagger\right),
	\end{aligned}
	\label{eq:full}
\end{equation}
with \(\hat{\sigma}_+=|e\rangle\langle g|\) and \(\hat{\sigma}_-=|g\rangle\langle e|\). The excited state \(|e\rangle\) decays at rate \(\Gamma\), and the density matrix obeys the Lindblad master equation
\begin{equation}
	\dot{\hat{\rho}}=-i[\hat{H}_{\mathrm{full}},\hat{\rho}]
	+\Gamma\left(\hat{\sigma}_-\hat{\rho}\hat{\sigma}_+
	-\frac{1}{2}\{\hat{\sigma}_+\hat{\sigma}_-,\hat{\rho}\}\right).
	\label{eq:lindblad}
\end{equation}
We define the bosonic collective operator
\begin{equation}
	\hat{s}=g_1e^{-i\phi_1}\hat{a}_1+g_2e^{-i\phi_2}\hat{a}_2,
	\label{eq:s}
\end{equation}
the internal-state matrix elements
\begin{equation}
	\hat{\rho}_{gg}=\langle g|\hat{\rho}|g\rangle,
	\qquad
	\hat{\rho}_{ee}=\langle e|\hat{\rho}|e\rangle,
	\qquad
	\hat{\rho}_{eg}=\langle e|\hat{\rho}|g\rangle,
	\qquad
	\hat{\rho}_{ge}=\langle g|\hat{\rho}|e\rangle,
	\label{eq:rhoij}
\end{equation}
and set
\begin{equation}
	\hat{H}_0=\hat{H}_{\mathrm{drive}}+\hat{H}_{\mathrm{coup}}.
	\label{eq:H0}
\end{equation}
The internal-state matrix elements satisfy
\begin{equation}
	\begin{aligned}
		\dot{\hat{\rho}}_{gg}
		&=-i[\hat{H}_0,\hat{\rho}_{gg}]
		-\frac{i}{2}\left(\hat{s}^\dagger\hat{\rho}_{eg}
		-\hat{\rho}_{ge}\hat{s}\right)
		+\Gamma\hat{\rho}_{ee},\\
		\dot{\hat{\rho}}_{ee}
		&=-i[\hat{H}_0,\hat{\rho}_{ee}]
		-\frac{i}{2}\left(\hat{s}\hat{\rho}_{ge}
		-\hat{\rho}_{eg}\hat{s}^\dagger\right)
		-\Gamma\hat{\rho}_{ee},\\
		\dot{\hat{\rho}}_{eg}
		&=-i[\hat{H}_0,\hat{\rho}_{eg}]
		-\frac{i}{2}\left(\hat{s}\hat{\rho}_{gg}
		-\hat{\rho}_{ee}\hat{s}\right)
		-\frac{\Gamma}{2}\hat{\rho}_{eg},\\
		\dot{\hat{\rho}}_{ge}
		&=-i[\hat{H}_0,\hat{\rho}_{ge}]
		-\frac{i}{2}\left(\hat{s}^\dagger\hat{\rho}_{ee}
		-\hat{\rho}_{gg}\hat{s}^\dagger\right)
		-\frac{\Gamma}{2}\hat{\rho}_{ge}.
	\end{aligned}
	\label{eq:internal}
\end{equation}
Under the condition \(\Gamma\gg g_1,g_2,\Omega,J\), the internal state relaxes much faster than the bosonic modes. Then, adiabatic elimination can be performed by setting \(\dot{\hat{\rho}}_{ee}\simeq\dot{\hat{\rho}}_{eg}\simeq\dot{\hat{\rho}}_{ge}\simeq0\) and neglecting the action of \(\hat{H}_0\) on these fast internal-state matrix elements. This gives
\begin{equation}
	\hat{\rho}_{ee}\simeq\frac{1}{2\Gamma}\hat{s}\hat{\rho}_{gg}\hat{s}^\dagger,
	\qquad
	\hat{\rho}_{eg}\simeq-\frac{i}{\Gamma}\hat{s}\hat{\rho}_{gg},
	\qquad
	\hat{\rho}_{ge}\simeq\frac{i}{\Gamma}\hat{\rho}_{gg}\hat{s}^\dagger.
	\label{eq:adiabatic}
\end{equation}
Substituting Eq.~\eqref{eq:adiabatic} into the equation for \(\hat{\rho}_{gg}\) in Eq.~\eqref{eq:internal} yields the effective master equation
\begin{equation}
	\dot{\hat{\rho}}=-i[\hat{H}_{\mathrm{eff}},\hat{\rho}]
	+\frac{1}{\Gamma}\mathcal{D}[\hat{s}]\hat{\rho},
	\label{eq:effmaster}
\end{equation}
where
\begin{equation}
	\hat{H}_{\mathrm{eff}}=\hat{H}_{\mathrm{drive}}+\hat{H}_{\mathrm{coup}},
	\qquad
	\mathcal{D}[\hat{s}]\hat{\rho}
	=\hat{s}\hat{\rho}\hat{s}^\dagger
	-\frac{1}{2}\{\hat{s}^\dagger\hat{s},\hat{\rho}\}.
	\label{eq:D}
\end{equation}
For notational simplicity, the reduced density matrix \(\hat{\rho}_{gg}\) in the effective master equation above has been relabeled as \(\hat{\rho}\).

\section {Mean-field analysis: nonreciprocity, battery steady state, and timescale of unidirectional charging}

Let \(\alpha_C=\langle\hat{a}_1\rangle\) and \(\alpha_B=\langle\hat{a}_2\rangle\). From Eq.~\eqref{eq:effmaster}, the mean amplitudes obey
\begin{equation}
	\begin{aligned}
		\frac{d\alpha_C}{dt}
		&=-i\left(\frac{J}{2}e^{-i\phi_c}-iGe^{i\phi_r}\right)\alpha_B
		-\gamma_C\alpha_C-i\Omega e^{i\phi_d},\\
		\frac{d\alpha_B}{dt}
		&=-i\left(\frac{J}{2}e^{i\phi_c}-iGe^{-i\phi_r}\right)\alpha_C
		-\gamma_B\alpha_B,
	\end{aligned}
	\label{eq:meanfield}
\end{equation}
where
\begin{equation}
	\gamma_C=\frac{g_1^2}{2\Gamma},
	\qquad
	\gamma_B=\frac{g_2^2}{2\Gamma},
	\qquad
	G=\frac{g_1g_2}{2\Gamma},
	\qquad
	\phi_r=\phi_1-\phi_2.
	\label{eq:gammas}
\end{equation}
Nonreciprocity requires that the charger receives no feedback from the battery, i.e., the coefficient of \(\alpha_B\) in the first equation of Eq.~\eqref{eq:meanfield} vanishes, which yields
\begin{equation}
	G=\frac{J}{2},
	\qquad
	\phi_c+\phi_r=-\frac{\pi}{2}.
	\label{eq:nonrecip}
\end{equation}
Under this condition, the dynamics becomes unidirectional:
\begin{equation}
	\frac{d\alpha_C}{dt}=-\gamma_C\alpha_C-i\Omega e^{i\phi_d},
	\qquad
	\frac{d\alpha_B}{dt}=-iJe^{i\phi_c}\alpha_C-\gamma_B\alpha_B.
	\label{eq:unidir}
\end{equation}
Setting \(d\alpha_C/dt=d\alpha_B/dt=0\) gives the steady-state amplitudes
\begin{equation}
	\alpha_C^{\mathrm{ss}}=-\frac{i\Omega}{\gamma_C}e^{i\phi_d},
	\qquad
	\alpha_B^{\mathrm{ss}}
	=-\frac{J\Omega}{\gamma_C\gamma_B}e^{i(\phi_c+\phi_d)}.
	\label{eq:steady1}
\end{equation}
Since \(G=J/2\) and \(G^2=\gamma_C\gamma_B\), we have \(\gamma_C\gamma_B=J^2/4\). Therefore
\begin{equation}
	\alpha_B^{\mathrm{ss}}
	=-\frac{4\Omega}{J}e^{i(\phi_c+\phi_d)}.
	\label{eq:steadyB}
\end{equation}
Eq.~\eqref{eq:steadyB} shows that, under the nonreciprocal condition, the steady-state battery amplitude depends only on \(\Omega\) and \(J\). Since Eq.~\eqref{eq:effmaster} is quadratic, contains no squeezing terms, and has a Lindblad operator \(\hat{s}\) composed of annihilation operators, Gaussianity is preserved. Thus, for the charging process starting from the vacuum \(\alpha_C(0)=\alpha_B(0)=0\), the modes remain in coherent states \(|\alpha_C(t)\rangle\otimes|\alpha_B(t)\rangle\) with amplitudes obeying Eq.~\eqref{eq:meanfield}, so the battery steady state is a coherent state \(|\alpha_B^{\mathrm{ss}}\rangle\).

For comparison, we now consider the purely coherent coupling case, obtained by setting \(g_1=g_2=0\), so that \(\gamma_C=\gamma_B=G=0\). The mean-field equations \eqref{eq:meanfield} then reduce to
\begin{equation}
	\frac{d\alpha_C}{dt}
	=-i\frac{J}{2}e^{-i\phi_c}\alpha_B-i\Omega e^{i\phi_d},
	\qquad
	\frac{d\alpha_B}{dt}
	=-i\frac{J}{2}e^{i\phi_c}\alpha_C.
	\label{eq:coherent}
\end{equation}
Starting again from the vacuum, \(\alpha_C(0)=\alpha_B(0)=0\), one finds
\begin{equation}
	\alpha_C(t)
	=-\frac{2i\Omega}{J}e^{i\phi_d}\sin\left(\frac{Jt}{2}\right),
	\qquad
	\alpha_B(t)
	=-\frac{2\Omega}{J}e^{i(\phi_c+\phi_d)}
	\left[1-\cos\left(\frac{Jt}{2}\right)\right].
	\label{eq:coherentsol}
\end{equation}
The battery amplitude oscillates and reaches its maximum at \(Jt/2=\pi\), where
\begin{equation}
	\alpha_B\left(t=\frac{2\pi}{J}\right)
	=-\frac{4\Omega}{J}e^{i(\phi_c+\phi_d)}
	=\alpha_B^{\mathrm{ss}}.
	\label{eq:maxamp}
\end{equation}
Thus, the maximum battery amplitude in the purely coherent coupling case exactly equals the steady-state amplitude obtained in the nonreciprocal dissipative case, including its phase. The difference lies in the dynamics: the purely coherent coupling exchanges energy periodically between the two modes, whereas the interference between coherent and dissipative couplings produces unidirectional charging that brings the battery monotonically to this oscillation peak value.

To analyze the charging timescale, we write the unidirectional mean-field equations \eqref{eq:unidir} in matrix form,
\begin{equation}
	\frac{d}{dt}
	\begin{pmatrix}
		\alpha_C\\ \alpha_B
	\end{pmatrix}
	=
	M
	\begin{pmatrix}
		\alpha_C\\ \alpha_B
	\end{pmatrix}
	+
	\begin{pmatrix}
		-i\Omega e^{i\phi_d}\\ 0
	\end{pmatrix},
	\qquad
	M=
	\begin{pmatrix}
		-\gamma_C & 0\\[4pt]
		-iJe^{i\phi_c} & -\gamma_B
	\end{pmatrix}.
	\label{eq:matrix}
\end{equation}
The eigenvalues of \(M\) are \(\lambda_1=-\gamma_C\) and \(\lambda_2=-\gamma_B\), so the transient dynamics toward the steady state is governed by these two rates.

For \(\gamma_C\neq\gamma_B\), the two eigenvalues are distinct, and the battery amplitude evolves as
\begin{equation}
	\alpha_B(t)=\alpha_B^{\mathrm{ss}}\left[
	1-\frac{\gamma_B}{\gamma_B-\gamma_C}e^{-\gamma_C t}
	+\frac{\gamma_C}{\gamma_B-\gamma_C}e^{-\gamma_B t}
	\right].
	\label{eq:neq}
\end{equation}
For \(\gamma_C=\gamma_B=\gamma\), the eigenvalues become degenerate, \(M\) is defective, and the system is at an exceptional point. The battery amplitude then evolves as
\begin{equation}
	\alpha_B(t)=\alpha_B^{\mathrm{ss}}\left[
	1-e^{-\gamma t}-\gamma t e^{-\gamma t}
	\right].
	\label{eq:eq}
\end{equation}
The \(t e^{-\gamma t}\) term is the signature of the exceptional point.

Under the nonreciprocal condition, the product \(\gamma_C\gamma_B=G^2=J^2/4\) is fixed, while the charging time is limited by the slower of the two rates, \(\min(\gamma_C,\gamma_B)\). This slower rate is maximized when \(\gamma_C=\gamma_B=J/2\). The exceptional point therefore corresponds to critical damping and yields the fastest approach to the battery steady state.

\section{Numerical simulation of unidirectional charging}

To verify the unidirectional charging dynamics and the accuracy of the adiabatic elimination, we perform numerical simulations of the full master equation. We use the experimental parameters of Fig.~3a in the main text: \(g_1=2\pi\times5\,\mathrm{kHz}\), \(g_2=2\pi\times6\,\mathrm{kHz}\), \(\Gamma=2\pi\times64\,\mathrm{kHz}\), \(J=2\pi\times0.469\,\mathrm{kHz}\), \(\Omega=2\pi\times0.225\,\mathrm{kHz}\), and \(\phi_c=-\pi/2\). Both motional modes are initialized in the vacuum, and the ion is initialized in its ground state.

\begin{figure*}[htbp]
	\centering
	\includegraphics*[scale=0.85]{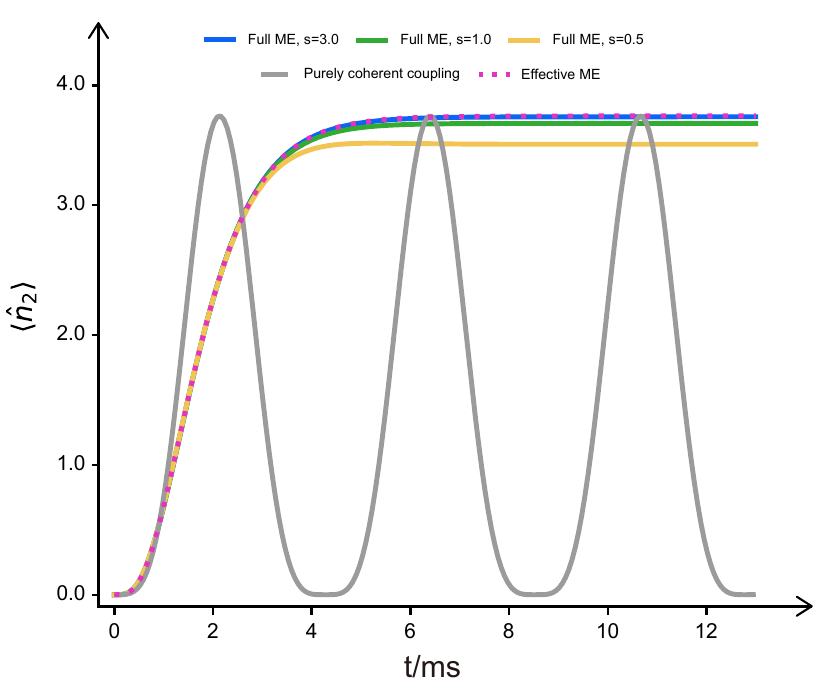}
	\caption{Time evolution of the battery occupation \(\langle \hat{n}_2\rangle\) during unidirectional charging. Results from the full master equation for different scaling factors \(s\) are compared with that from the effective master equation after adiabatic elimination. The result for the purely coherent coupling case is also shown for comparison. The parameters are the same as in Fig.~3a of the main text.
	}
	\label{figureS1}
\end{figure*}

Fig.~\ref{figureS1} shows the time evolution of the battery occupation \(\langle \hat n_2\rangle\), where \(\hat n_2=\hat a_2^\dagger \hat a_2\). The colored solid curves are obtained from the full master equation. The gray solid curve corresponds to the purely coherent coupling case with \(g_1=g_2=0\),  which exhibits periodic oscillations. The magenta dotted curve is obtained from the effective master equation after adiabatic elimination. As seen in Fig.~\ref{figureS1}, the full master equation approaches a steady-state value slightly below the coherent oscillation peak
\begin{equation}
	|\alpha_B|_{\max}^2 = \left(\frac{4\Omega}{J}\right)^2 \simeq 3.686.
	\label{eq:peak}
\end{equation}
This deviation arises from the imperfect adiabatic approximation due to the finite ratio \(g_i/\Gamma\).

To systematically quantify this effect, we introduce a scaling factor \(s\) such that
\begin{equation}
	g_1 \to s\,g_1,\qquad
	g_2 \to s\,g_2,\qquad
	\Gamma \to s^2\,\Gamma.
	\label{eq:scaling}
\end{equation}
Under this scaling, all effective parameters remain invariant,
\begin{equation}
	\gamma_C=\frac{g_1^2}{2\Gamma},\qquad
	\gamma_B=\frac{g_2^2}{2\Gamma},\qquad
	G=\frac{g_1g_2}{2\Gamma},\qquad
	J=\frac{g_1g_2}{\Gamma},
	\label{eq:invariant}
\end{equation}
so that the effective master equation is unchanged. The coherent oscillation peak, \(|\alpha_B|_{\max}^2=(4\Omega/J)^2\), is also unchanged, since both \(J\) and \(\Omega\) are unaffected by \(s\). However, the ratio \(g_i/\Gamma\) scales as \(1/s\), so the adiabatic approximation becomes more accurate as \(s\) increases. Fig.~\ref{figureS1} compares the battery occupation for different values of \(s\). As \(s\) increases, the steady-state value of \(\langle \hat n_2\rangle\) approaches the coherent oscillation peak \(|\alpha_B|_{\max}^2\simeq 3.686\). This confirms that the deviation between the full master equation and the effective master equation is determined by the ratio \(g_i/\Gamma\), and that the unidirectional charging indeed reaches the coherent oscillation peak in the ideal adiabatic limit.

\section{Effects of detuning on the dynamics}
The theoretical analysis in the preceding sections assumes that all drives and couplings are perfectly resonant. In practice, however, experimental imperfections inevitably introduce small detunings. It is therefore important to analyze how these detunings affect the dynamics, to assess the requirements on experimental control, and to understand the observed deviations from the ideal predictions.

We consider three types of detuning: drive detuning \(\delta_1\), which shifts the frequency of the external drive applied to the charger mode; coupling detuning \(\delta_2\), which shifts the frequency of the coherent coupling between the two motional modes; and red-sideband (RSB) detuning \(\delta_3\), which shifts the frequency of the Jaynes--Cummings coupling between the ion's internal state and the charger mode.  Adding these detunings to the resonant Hamiltonian of Eq.~\eqref{eq:full} gives
\begin{equation}
	\begin{aligned}
		\hat H(t)
		&=
		\Omega\left(\hat a_1^\dagger+\hat a_1\right)
		+\delta_1\,\hat a_1^\dagger\hat a_1
		\\
		&\quad+\frac{J}{2}
		\left(
		e^{-i\phi_c}e^{i\delta_2 t}\hat a_1^\dagger\hat a_2
		+
		e^{i\phi_c}e^{-i\delta_2 t}\hat a_1\hat a_2^\dagger
		\right)
		\\
		&\quad+\frac{g_1}{2}
		\left(
		e^{-i\phi_1}e^{i\delta_3 t}\hat\sigma_+\hat a_1
		+
		e^{i\phi_1}e^{-i\delta_3 t}\hat\sigma_-\hat a_1^\dagger
		\right)
		\\
		&\quad+\frac{g_2}{2}
		\left(
		e^{-i\phi_2}\hat\sigma_+\hat a_2
		+
		e^{i\phi_2}\hat\sigma_-\hat a_2^\dagger
		\right),
	\end{aligned}
	\label{eq:H_det}
\end{equation}
where \(J=g_1g_2/\Gamma\) is fixed by the nonreciprocal condition \(G=J/2\). The dissipative term remains unchanged. 

\begin{figure*}[htbp]
	\centering
	\includegraphics*[scale=0.85]{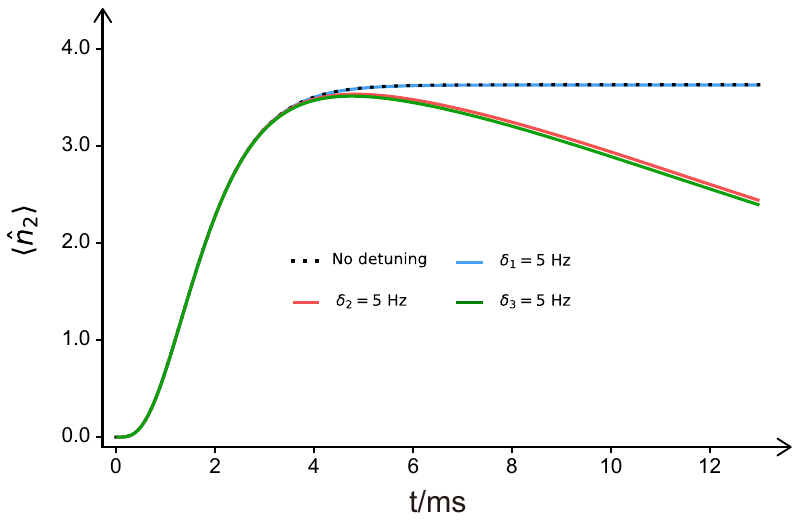}
	\caption{Battery occupation \(\langle \hat n_2\rangle\) in the nonreciprocal charging process, comparing the resonant case with three individual \(5\) Hz detunings \(\delta_1\), \(\delta_2\), and \(\delta_3\), each applied separately. Parameters are the same as in Fig.~\ref{figureS1} with \(s=1\). 
	}
	\label{figureS2}
\end{figure*}

We first study the effect of detuning on the nonreciprocal charging process. Fig.~\ref{figureS2} shows the time evolution of the battery occupation \(\langle \hat n_2\rangle\) obtained from numerical simulations, comparing the resonant case with three cases where a single detuning among \(\delta_1,\delta_2,\delta_3\) is set to \(5\) Hz while the other two are zero. The drive detuning \(\delta_1\) has a negligible effect, while the coupling detuning \(\delta_2\) and the RSB detuning \(\delta_3\) both cause the occupation to peak near \(4\) ms and then decrease, resulting in a lower steady-state value.

\begin{figure*}[htbp]
	\centering
	\includegraphics*[scale=0.85]{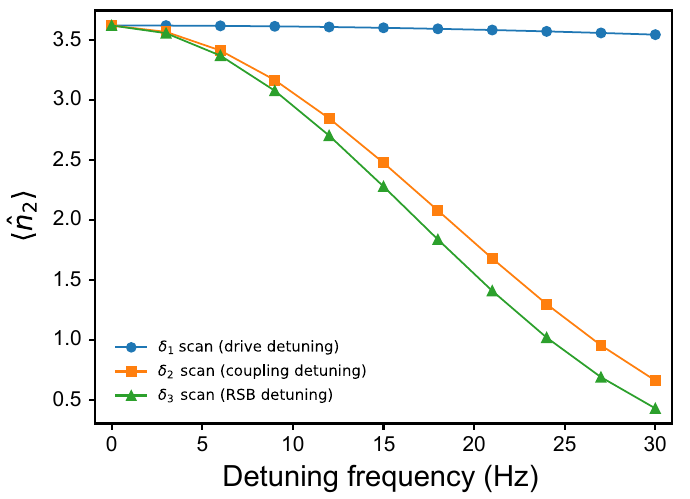}
	\caption{Battery occupation \(\langle \hat n_2\rangle\) at \(t=6\) ms as a function of the detuning frequency. Each curve is obtained by scanning one of the detunings \(\delta_1\), \(\delta_2\), or \(\delta_3\), while the other two are set to zero. Other parameters are the same as in Fig.~\ref{figureS1} with \(s=1\).
	}
	\label{figureS3}
\end{figure*}

To quantify the sensitivity to each detuning, we fix the evolution time at \(t=6\) ms and scan the detuning frequency from \(0\) to \(30\) Hz. The resulting \(\langle \hat n_2\rangle\) is plotted in Fig.~\ref{figureS3}. The drive detuning \(\delta_1\) has a negligible effect: even at \(20\) Hz, the occupation remains close to the resonant value. In contrast, both the coupling detuning \(\delta_2\) and the RSB detuning \(\delta_3\) cause a rapid decrease in \(\langle \hat n_2\rangle\), with \(\delta_3\) being slightly more detrimental than \(\delta_2\). To maintain acceptable charging performance, these two detunings must be controlled within about \(5\) Hz. This imposes stringent requirements on the stability of the motional frequencies in the trapped-ion experiment.

\begin{figure*}[bhtp]
	\centering
	\includegraphics*[scale=0.78]{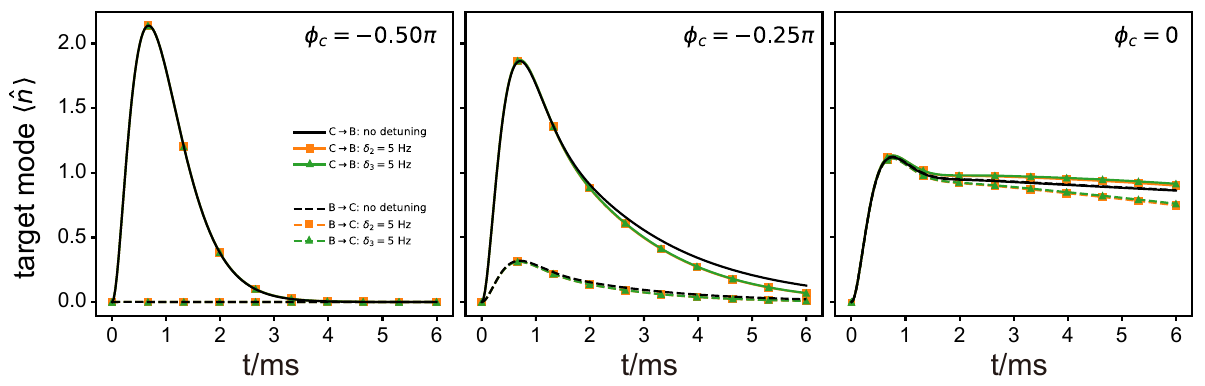}
	\caption{Energy transfer without external drive for two initial conditions: a coherent state \(|\alpha=2\rangle\) is initially stored in either the charger (solid) or the battery (dashed), while the target mode is in the vacuum. For each initial condition, the resonant case is compared with two detuned cases, where only one of \(\delta_2,\delta_3\) is set to \(5\) Hz and the other is zero. Panels correspond to \(\phi_c=-\pi/2\), \(-\pi/4\), and \(0\). Other parameters are the same as in Fig.~\ref{figureS1} with \(s=1\).
	}
	\label{figureS4}
\end{figure*}

We now turn to the second scenario, where the external drive is absent. Fig.~\ref{figureS4} shows the results for two initial conditions: the energy is initially stored either in the charger mode or in the battery mode, prepared in a coherent state with \(\alpha=2\), while the other mode is in the vacuum. For each initial condition, we compare the resonant case with the cases where only the coupling detuning \(\delta_2\) or the RSB detuning \(\delta_3\) is set to \(5\) Hz. This allows us to analyze how detuning affects the nonreciprocal energy transfer shown in Fig.~2 of the main text. For the optimal phase \(\phi_c=-\pi/2\), the energy transfer is less sensitive to detuning than for the other two phases. For \(\phi_c=-\pi/4\) and \(0\), however, the detuning must be controlled within about \(5\) Hz to keep the evolution close to the resonant case. Since the experimental and theoretical curves agree well, the frequency instability in our experiment is less than \(5\) Hz.

\section{Experimental Apparatus}
\label{sec:apparatus}

\begin{figure}[t]
	\centering
	\includegraphics[width=0.8\textwidth]{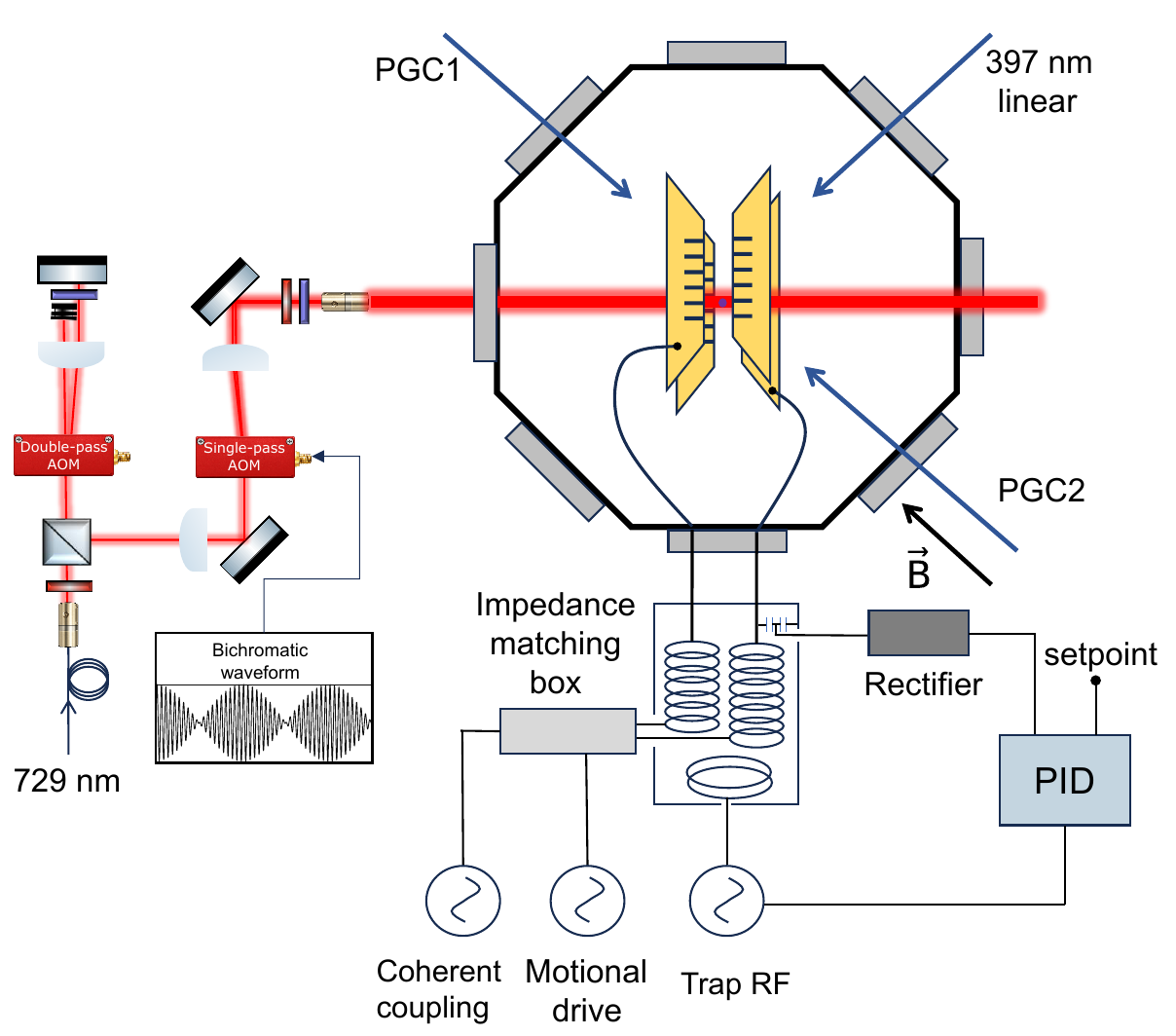}
	\caption{
		Schematic of the experimental apparatus, including the electric motional-drive circuit, the optical system for generating the bichromatic field, laser cooling of the ion, and the trap-frequency PID stabilization module.
	}
	\label{FigS5}
\end{figure}

The experimental schematic is illustrated in Fig.~\ref{FigS5}. A single $^{40}\text{Ca}^+$ ion is confined in a segmented blade Paul trap, featuring radial secular frequencies of approximately $2\pi \times \{2.71, 2.38\}\text{ MHz}$ (marked as mode B and C) and an axial frequency of  $\sim 2\pi \times 500\text{ kHz}$. For the two radial modes, the heating rates are measured to be $\dot{\bar{n}} = 5\text{ quanta/s}$ and $\dot{\bar{n}} = 18\text{ quanta/s}$, with corresponding motional coherence times exceeding $6\text{ ms}$ and $11\text{ ms}$, respectively. To suppress drift in the radial trap frequencies, an active stabilization network is implemented: the RF voltage at the trap blade is sampled via a voltage divider, rectified into a DC signal, and routed to a proportional-integral-derivative (PID) controller. The PID loop then feeds back to the amplitude modulation port of the RF signal generator, ensuring long-term trap stability. The charger mode is driven by applying an RF field to a single trap blade at a frequency resonant with the higher-frequency radial mode. Coherent coupling between the two radial modes is achieved by simultaneously driving diagonal blades with an RF field resonant at the frequency difference between the modes. To integrate the micromotion compensation and motional modulation signals, a custom impedance matching network was fabricated following the approach outlined in Ref.~\cite{Metzner2024}.

\begin{figure}[t]
	\centering
	\includegraphics[width=0.8\textwidth]{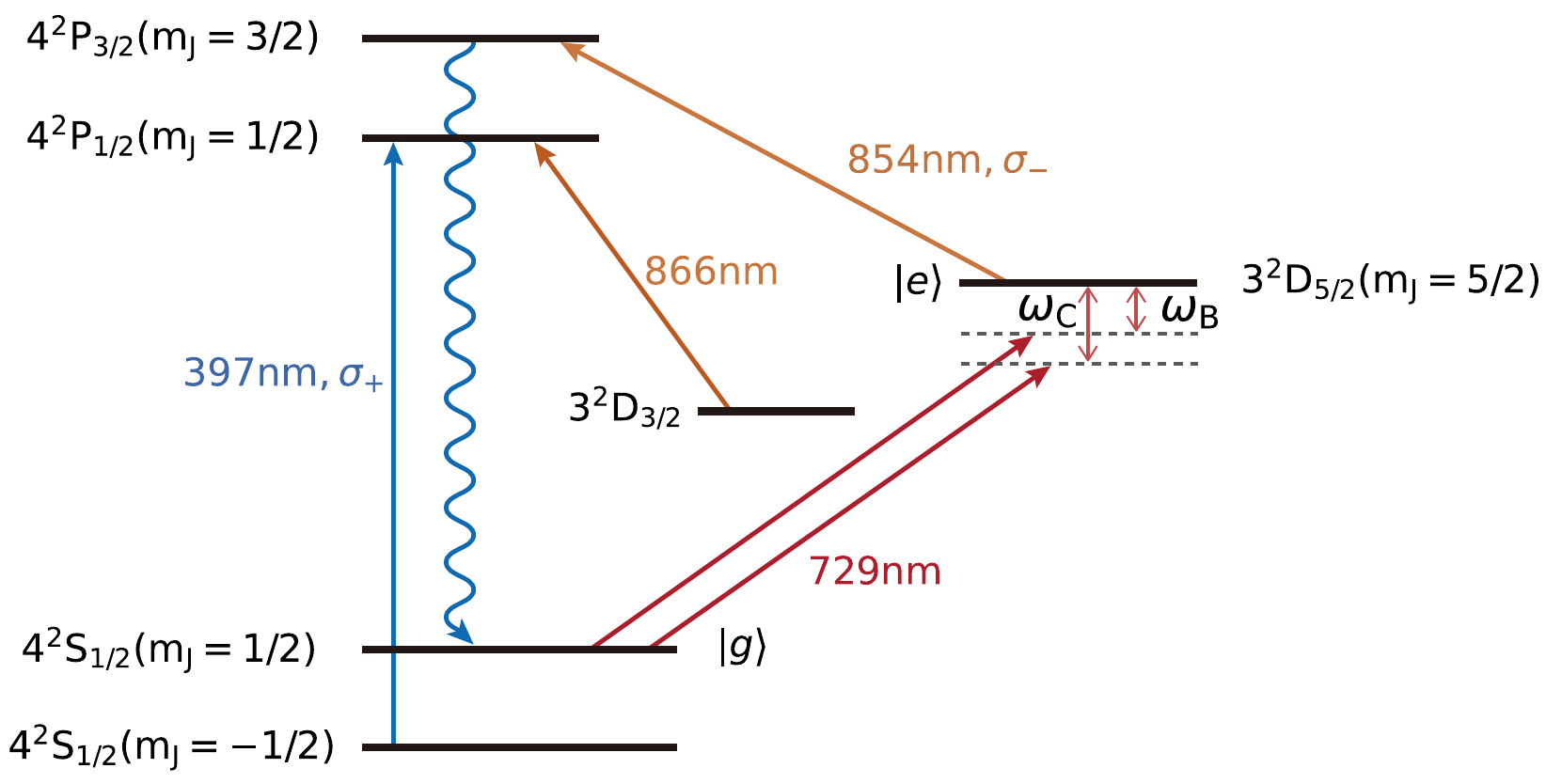}
	\caption{
		Relevant energy levels and optical transitions of $^{40}\mathrm{Ca}^{+}$ used in the experiment.
	}
	\label{FigS6}
\end{figure}

A static magnetic field of $\sim 7.35\text{ G}$ is applied at an angle of approximately $45^\circ$ relative to the trap axis to define the quantization axis. The relevant energy level structure of $^{40}\text{Ca}^+$ is depicted in Fig.~\ref{FigS6}. The optical qubit is encoded in the ground and metastable states, designated as $\lvert g\rangle \equiv \lvert 4s\text{ }^2S_{1/2}, m_J = 1/2\rangle$ and $\lvert e\rangle \equiv \lvert 3d\text{ }^2D_{5/2}, m_J = 5/2\rangle$. Coherent driving of the narrow optical quadrupole transition at $729\text{ nm}$ is realized by an ultra-stable laser system with a linewidth of $\sim 10\text{ Hz}$, yielding a qubit coherence time exceeding $4\text{ ms}$ without fiber noise cancellation.

Frequency translation and pulse shaping of the $729\text{ nm}$ beam are managed by a network of acousto-optic modulators (AOMs). A double-pass AOM provides coarse frequency tuning, while a subsequent single-pass AOM, driven by a multi-tone RF signal, modulates the beam to simultaneously address the two RSBs of the radial modes. The $729\text{ nm}$ laser propagates perpendicular to the trap axis, with the corresponding Lamb-Dicke parameters evaluated as $0.0505$ and $0.0378$ for the respective radial modes. Additionally, an $854\text{ nm}$ laser beam, co-propagating with the $397\text{ nm}$ detection beam, is employed to realize a dissipation channel with a tunable rate $\Gamma$, and to reset the qubit state at the end of each experimental sequence.

Each experimental cycle begins with Doppler cooling on the $\lvert 4s\text{ }^2S_{1/2}\rangle \leftrightarrow \lvert 4p\text{ }^2P_{1/2}\rangle$ transition using a red-detuned $397\text{ nm}$ laser beam, while an $866\text{ nm}$ repumping laser simultaneously clears out population from the metastable $\lvert 3d\text{ }^2D_{3/2}\rangle$ state. Subsequently, sub-Doppler cooling is performed via polarization gradient cooling (PGC) using two counter-propagating, linearly polarized $397\text{ nm}$ beams aligned along the quantization axis~\cite{JoshiRoos2020,LiSchmidt-Kaler2022,EjtemaeeHaljan2017}, cooling all motional modes to a mean phonon number of $\bar{n} \approx 2$. 

Further cooling to the motional ground state is accomplished via resolved sideband cooling, achieving a  mean phonon occupation of $\bar{n} < 0.1$ for both radial modes. Prior to executing quantum operations, the internal state is initialized into the state $\lvert 4s\text{ }^2S_{1/2}, m_J = 1/2\rangle$ via optical pumping. For state readout, a $397\text{ nm}$ probe beam resonant with the $\lvert 4s\text{ }^2S_{1/2}\rangle \leftrightarrow \lvert 4p\text{ }^2P_{1/2}\rangle$ transition is applied. The resulting ion fluorescence is collected by a high-numerical-aperture objective ($\text{NA} = 0.4$) and imaged onto a CCD camera for state discrimination.

\section{Experimental parameter calibration}

The quantum battery experiment imposes stringent requirements on the precision of experimental parameters, including the trap frequency, dissipation rate, and amplitudes of the motional drive and coupling. Consequently, all these parameters must be carefully calibrated prior to data acquisition. This section describes the calibration procedures for these parameters.

\begin{figure}[t]
	\centering
	\includegraphics[width=1.0\textwidth]{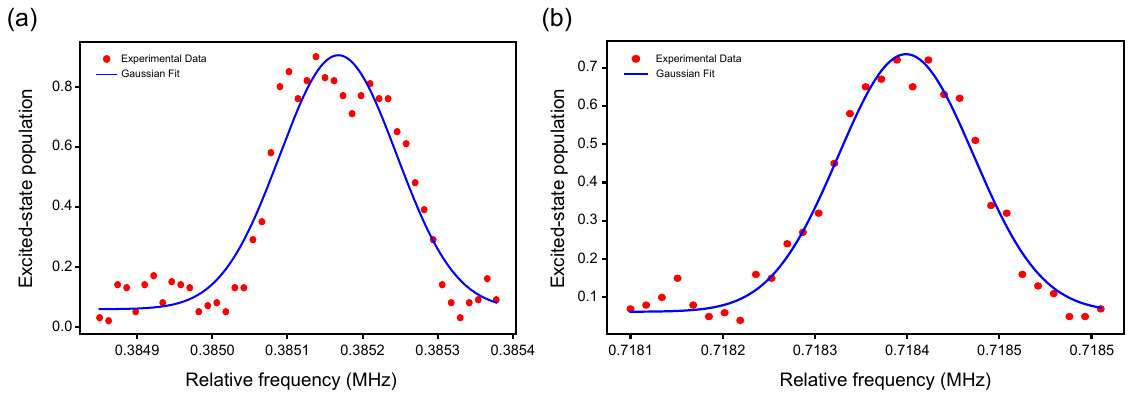}
	\caption{Calibration of the trap frequency using a tickle pulse and red-sideband spectrum. (a) and (b) show calibration examples for modes B and C, respectively. Note that the horizontal axis denotes the frequency relative to 2 MHz.}
	
	\label{FigS7}
\end{figure}

The trap frequency stability is critical for both the motional drive and coherent coupling operations; theoretical simulations indicate that the long-term frequency drift and the measurement uncertainty for each motional mode should be less than 5Hz during the experiment. To maintain the trap frequency stability, we built a trap-voltage sampling and locking circuit following the method in \cite{JohnsonMonroe2016, ZhangHaljan2022}, but the stability remained insufficient because the electrical components in the locking loop are sensitive to laboratory temperature drift. This problem was solved by employing a temperature control box for the entire vacuum chamber and the trap frequency locking loop, which reduces the temperature fluctuation to below $0.1$ degree. The precise trap frequency measurement is achieved by monitoring the mode's response to a weak probe RF pulse applied to one of the trap blades (the tickle method). The pulse duration is fixed at $5$ ms, and its frequency is scanned. When the pulse frequency resonates with the motional mode, the ion initially in the ground state is driven into a coherent state with maximum amplitude, and the RSB transition is maximally excited. The precise trap frequency is then obtained by fitting the RSB excitation spectrum. Fig.~\ref{FigS7} shows the trap frequency measurement results for modes B and C using the tickle method.

\begin{figure}[t]
	\centering
	\includegraphics[width=0.6\textwidth]{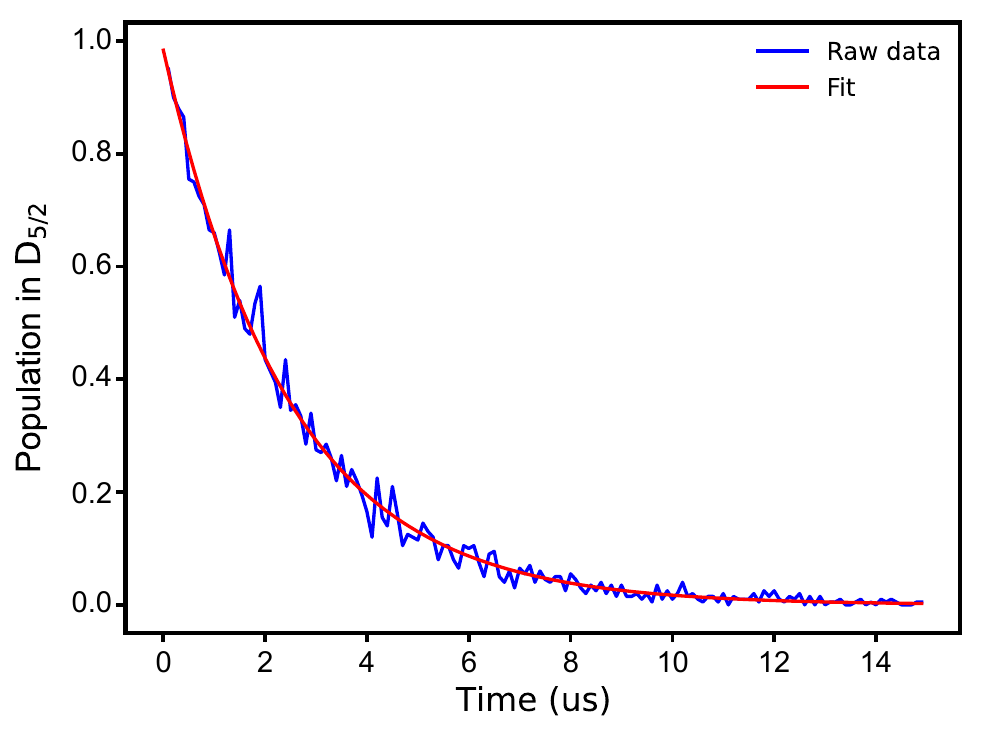}
	\caption{
		Calibration of the dissipation rate induced by the 854 nm laser beam. The blue curve connects the raw experimental data points, and the red curve is an exponential fit.
	}
	\label{FigS8}
\end{figure}

Controllable dissipation of the $|e\rangle$ state is essential for realizing the dissipative coupling, which is tuned by adjusting the power of the $854$ nm laser beam. We calibrate the $854$ nm laser power by measuring the dissipation rate from $D_{5/2}(m_J=5/2)$ to $S_{1/2}(m_J=1/2)$. In the calibration procedure, a $729$ nm $\pi$ pulse is first applied to prepare the ion in the $D_{5/2}(m_J=5/2)$ state. An $854$ nm pulse resonant with the ${3}^{2}D_{5/2}(m_J=5/2) \leftrightarrow {4}^{2}P_{3/2}(m_J=3/2)$ transition is then used to pump the population out of $D_{5/2}(m_J=5/2)$, after which the remaining population in $D_{5/2}(m_J=5/2)$ is detected. Finally, by performing an exponential fit of the $D_{5/2}$ population versus the duration of the $854$ nm pulse, the dissipation rate is extracted. Fig.~\ref{FigS8} shows the fitted dissipation rate of $2\pi \times 64$ kHz, with an uncertainty of less than $2\pi \times 1$ kHz.

\begin{figure}[t]
	\centering
	\includegraphics[width=0.9\textwidth]{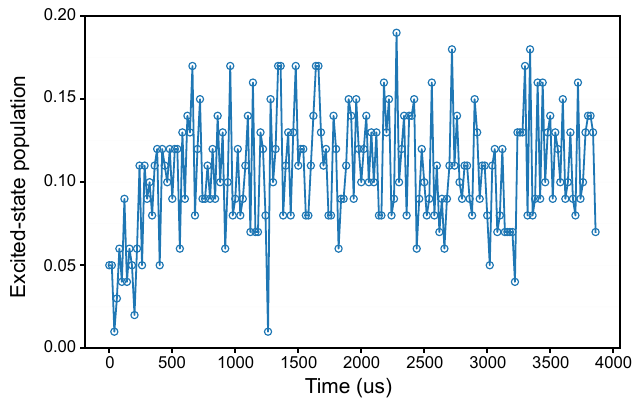}
	\caption{Dissipative coupling is verified by monitoring the phonon dynamics of mode B via its red-sideband excitation probability. Initially, mode B is prepared in the ground state, while mode C is in a thermal state with a mean phonon number of approximately 2. The dissipative-coupling laser pulse consists of a bichromatic 729 nm beam and an 854 nm beam.}
	\label{FigS9}
\end{figure}

By precisely setting the trap frequencies, the dissipative coupling between the two radial modes is realized through a two-mode JC model with a shared dissipative two-level system. Experimentally, the successful implementation of this coupling mechanism can be verified by the ability to prepare dissipation-engineered steady states. Specifically, a bichromatic $729$ nm laser beam with two distinct frequency components is applied to simultaneously drive the red sidebands of the two radial motional modes, thereby realizing the two-mode JC coupling. In the $N$-phonon subspace, the two-mode JC Hamiltonian yields maximally bright and perfectly dark eigenstates \cite{Parke2024}. Concurrently, the $854$ nm-induced dissipation creates a continuous sideband cooling channel for the bright phononic state components, leaving the dark components unaffected. As a result, the bright-state population is continuously depleted, steering the system to asymptotically settle into the phononic steady state of the two radial motional modes. Note that the AC Stark shift induced by the two RSB transitions is not directly compensated; instead, we optimize the corresponding sideband cooling by searching for the appropriate frequency of the carrier transition and identify it as the shifted carrier frequency. In our experiment, we first cool the battery mode (mode B) to the ground state and prepare the charger mode (mode C) in a thermal state with a mean phonon number of approximately 2. The dissipative-coupling laser pulse is then turned on, and the phonon dynamics of mode B is monitored via its red-sideband transition probability. A steady state is observed, as shown in Fig.~\ref{FigS9}, confirming the successful implementation of the dissipative coupling mechanism.

\begin{figure}[t]
	\centering
	\includegraphics[width=0.9\textwidth]{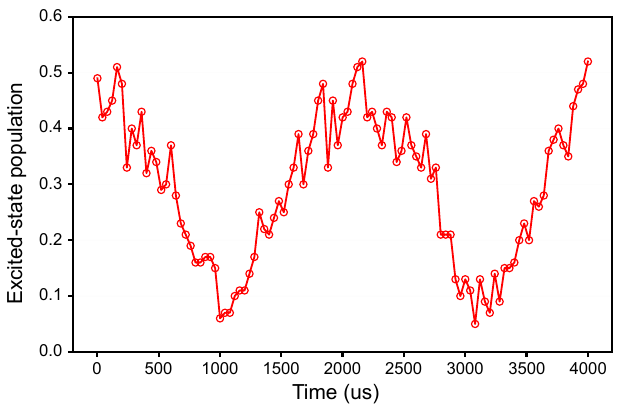}
	\caption{Phonon dynamics of mode C monitored via its red-sideband transition probability under coherent coupling. Mode C is initially prepared in a thermal state with a mean phonon number of approximately 2, while mode B is in the ground state. The oscillation period is inversely proportional to the coupling strength, which is approximately $2\pi\times0.47\;\mathrm{kHz}$.}
	
	\label{FigS10}
\end{figure}

The motional coherent coupling is realized by applying two identical RF signals to the diagonal trap blades, with the signal frequency tuned to equal the frequency difference between the two motional modes. The realization of coherent coupling is verified by observing the phonon swapping between the motional modes. In our calibration procedure, mode B is cooled to the ground state and mode C is prepared in a thermal state with an average phonon number of approximately 2, the coherent coupling strength $J$ is extracted by fitting the phonon swapping dynamics. As shown in Fig.~\ref{FigS10}, the phonon state in mode C exhibits periodic oscillatory behavior by monitoring its red-sideband transition probability. The minimum of the curve is nearly zero, indicating a near-perfect swap between the two modes. The oscillation period is inversely proportional to the coupling strength.

By combining the dissipative and coherent couplings, the nonreciprocal energy transfer between the two motional modes can be experimentally engineered by tailoring the quantum interference between these two channels. Initially, the coupling strengths are calibrated to be equal, with one motional mode initialized in the ground state and the other prepared in a coherent state $|\alpha\rangle$ with $|\alpha|=2$. Notably, the strength of the dissipative coupling scales as $g_1 g_2/(2\Gamma)$, where $g_i$ ($i=1,2$) represents the Rabi frequency of the RSB transition for mode $i$, and $\Gamma$ denotes the dissipation rate induced by the $854$ nm laser. Subsequently, the coherent coupling drive, the $854$ nm laser, and the two-tone $729$ nm laser beam are applied simultaneously. At the same time, the RF phase $\phi_c$ for the coherent coupling is systematically scanned to identify the optimal condition for destructive quantum interference. Under this optimized configuration, the coherent state initially in one motional mode transfers unidirectionally to the colder mode, as shown in Fig.~2b of the main text.

%
%

\section{Reconstruction of Fock state distribution from blue-sideband Rabi oscillations}
\label{sec:phonon}

\subsection{Theoretical Model and Optimization Protocol}
To reconstruct the Fock state population distribution $P(n)$ from the measured time-dependent excited-state populations $P_{\text{exp}}(t)$ during blue-sideband (BSB) Rabi oscillations, we employ a constrained nonlinear least-squares fitting framework based on the sequential least-squares programming (SLSQP) algorithm. The coherent dynamics of the single-ion BSB transition are governed by the model function:

\begin{equation}
	P_{\text{BSB}}(t) = \frac{1}{2} \sum_{n=0}^{n_{\text{max}}} P(n) \left[ 1 - e^{-\gamma t} \cos\!\left(\sqrt{n+1}\, g\, t\right) \right],
	\label{eq:s1}
\end{equation}
where $P_{\text{BSB}}(t)$ is the theoretical excited-state population for the BSB dynamics, and the Fock space is truncated at a physically justified limit of $n_{\text{max}} = 10$. The exponential damping coefficient $\gamma$ accounts for ambient motional dephasing, as also used in Ref.~\cite{Lv2018}; the BSB Rabi frequency $g = \eta g_{c}$ is rigorously calibrated via independent measurements of the carrier Rabi frequency $g_{c}$ (with Lamb-Dicke parameter $\eta = 0.0378$).

An initial guess for $P(n)$ is constructed based on a displaced coherent state representation. The optimizer minimizes the objective loss function, defined as the residual sum of squares (RSS):
\begin{equation}
	\mathrm{RSS} = \sum_{t} \left[ P_{\mathrm{exp}}(t) - P_{\mathrm{BSB}}(t) \right]^2.
\end{equation}
To ensure that the reconstructed density matrix of the phonon state remains physically valid, strict bounds and a normalization constraint are enforced during optimization:
\begin{align}
	\text{Bounds: } & P(n) \ge 0, \quad \forall n \in [0, n_{\max}], \\
	\text{Constraint: } & \sum_{n=0}^{n_{\max}} P(n) = 1.
\end{align}

\begin{figure}[t]
	\centering
	\includegraphics[width=1\textwidth]{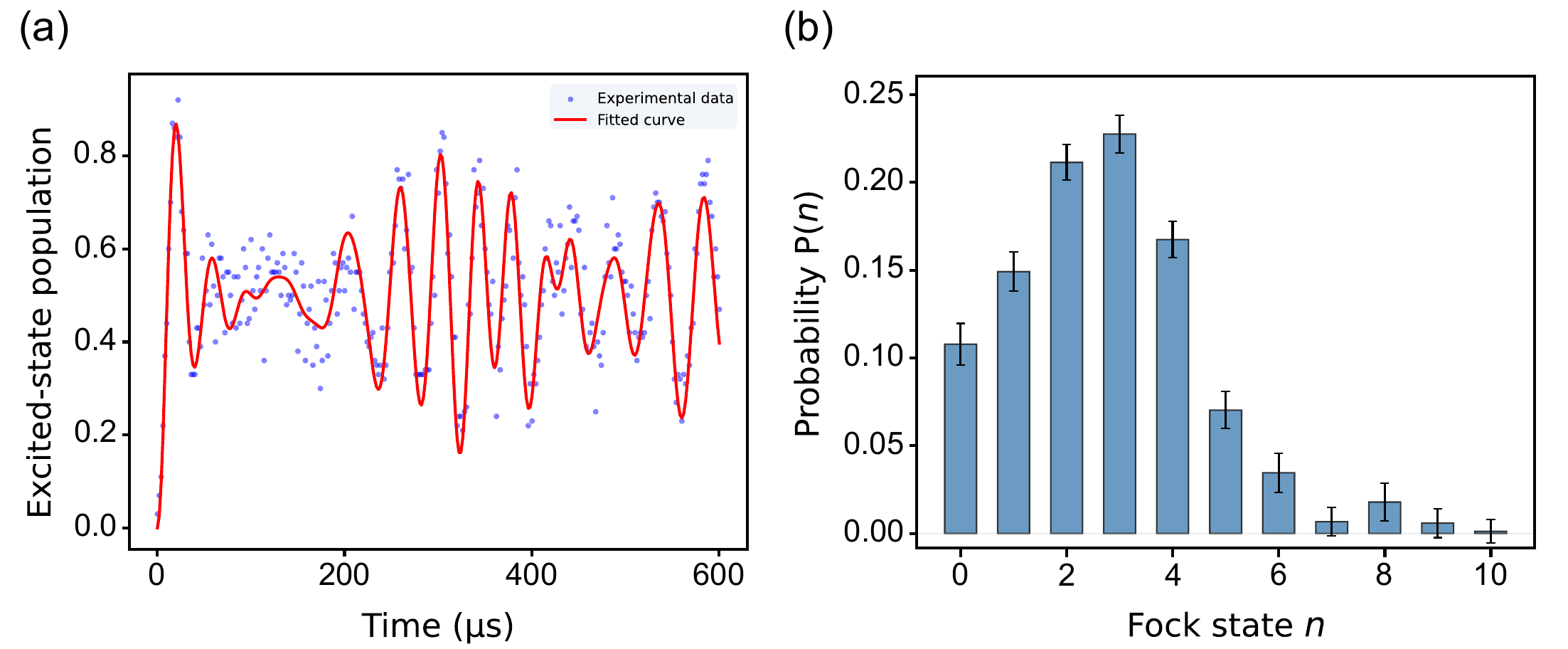}
	\caption{
		Experimental data and fitting results for the displaced coherent state in the quantum battery experiment. (a) Experimental data points are fitted using Eq.~\eqref{eq:s1}. (b) the resulting Fock-state probability distribution.
	}
	\label{FigS11}
\end{figure}

Once converged, the optimized probability distribution $P(n)_{\mathrm{fit}}$ is obtained, from which the mean phonon number $\langle n \rangle$ is directly computed as the expectation value
\begin{equation}
	\langle n \rangle = \sum_{n=0}^{n_{\mathrm{max}}} n \cdot P(n)_{\mathrm{fit}}.
\end{equation}
Fig.~\ref{FigS11} shows the fitted $P(n)$ for the displaced coherent state in Fig.~3c. The mean phonon number for this state is determined as $2.739 \pm 0.177$.

\subsection{Uncertainty Estimation via Bootstrap Resampling}
To obtain statistically robust uncertainty estimates without relying on Gaussian assumptions or being affected by local-minima artifacts, we use nonparametric bootstrap resampling.

The experimental dataset, consisting of paired time--probability data points $(t_i, P_i)$, is resampled with replacement to generate $B = 200$ independent synthetic datasets of identical size. For each bootstrap iteration, the full SLSQP optimization protocol is re-executed under the same boundary conditions and normalization constraint, yielding an ensemble of fitted distributions $\{P^{(b)}(n)\}_{b=1}^{B}$. The standard error $\Delta P(n)$ for each individual Fock-state probability is computed as the standard deviation of the bootstrap ensemble:
\begin{equation}
	\Delta P(n) = \sqrt{\frac{1}{B-1} \sum_{b=1}^{B} \bigl(P^{(b)}(n) - \bar{P}(n)\bigr)^2}.
\end{equation}
Finally, the standard errors $\Delta P(n)$ are propagated in quadrature to obtain the total uncertainty $\Delta\langle n\rangle$ of the mean phonon number:
\begin{equation}
	\Delta\langle n\rangle = \sqrt{\sum_{n=0}^{n_{\max}} \bigl( n\cdot\Delta P(n) \bigr)^2}.
\end{equation}


\section{Tomography of the Motional Wigner Function via Characteristic Function Measurements}
\label{sec:wigner_reconstruction}

This section details the phase-space reconstruction method used to obtain the motional Wigner function presented in Fig.~3. Rather than directly measuring the Wigner distribution, our approach samples the symmetrically ordered characteristic function, then applies a two-dimensional Fourier transform to reconstruct the Wigner function, as demonstrated in Ref.~\cite{FluhmannHome2020}. In the following, we outline the measurement protocol and the key techniques involved, including spin-dependent displacement, phase synchronization, characteristic-function sampling, and the Fourier transform protocol for obtaining the Wigner profile.

\subsection{Measurement Protocol and Phase Synchronization}
\label{Protocol}

In our trapped-ion experiment, the motional state is manipulated and read out via the auxiliary internal electronic levels of the ion, forming a spin system with basis states $\ket{g}$ and $\ket{e}$. The corresponding Pauli operators are defined as $\hat{\sigma}_x = \ket{e}\bra{g} + \ket{g}\bra{e}$, $\hat{\sigma}_y = -i\ket{e}\bra{g} + i\ket{g}\bra{e}$, and $\hat{\sigma}_z = \ket{e}\bra{e} - \ket{g}\bra{g}$, with eigenstates of $\hat{\sigma}_x$ given by $\ket{\pm} = (\ket{e} \pm \ket{g})/\sqrt{2}$. To evaluate a given motional quantum state $\hat{\rho}_{m}$, we sample its symmetrically ordered characteristic function, defined as
\begin{equation}
	\chi(\beta)=\operatorname{Tr}[\hat{\rho}_{m} \hat{D}(\beta)],
\end{equation}
where $\hat{D}(\beta)=\exp(\beta \hat{a}^{\dagger}-\beta^{*}\hat{a})$ is the displacement operator in phase space, $\hat{a}^{\dagger}$ and $\hat{a}$ are the creation and annihilation operators of the motional mode, and $\operatorname{Tr}$ denotes the trace operation.

The measurement protocol for $\chi(\beta)$ proceeds as follows  \cite{FluhmannHome2020}. First, the spin is initialized in $\ket{e}$. A resonant carrier rotation $\hat{R}(\theta,\phi)$ is then applied to the spin, parameterized by the rotation angle $\theta$ and the laser phase $\phi$:
\begin{equation}
	\hat{R}(\theta,\phi) = \cos(\theta/2)\,\hat{\mathbbm{I}} - i\sin(\theta/2)\bigl[\cos\phi\,\hat{\sigma}_x + \sin\phi\,\hat{\sigma}_y\bigr],
\end{equation}
where $\hat{\mathbbm{I}}$ is the identity operator in the spin subspace. Next, a bichromatic laser pulse generates a state-dependent displacement, described by $\hat{D}\bigl(\beta\,\hat{\sigma}_x/2\bigr)$, on the motional state, which displaces the $\ket{+}$ component by $+\beta/2$ and the $\ket{-}$ component by $-\beta/2$ in phase space. Finally, the real and imaginary parts of $\chi(\beta) = \langle \hat{D}(\beta) \rangle$ are extracted via projective measurements of the spin.

Specifically, with the laser phase set to $\phi = 0$ for the carrier rotation, the expectation value of the spin operator $\hat{\sigma}_z$ yields:
\begin{equation}
	\langle\hat{\sigma}_z\rangle = \frac12\bigl\langle e^{i\theta}\hat{D}(-\beta) + e^{-i\theta}\hat{D}(\beta)\bigr\rangle = \cos\theta\,\operatorname{Re}[\chi(\beta)] + \sin\theta\,\operatorname{Im}[\chi(\beta)].
\end{equation}
Thus, the real and imaginary parts of $\chi(\beta)$ can be extracted by choosing the carrier rotation angle $\theta$ appropriately:
\begin{align}
	\langle\hat{\sigma}_z\rangle &= \operatorname{Re}\,\chi(\beta) \quad (\text{for }\theta=0), \\
	\langle\hat{\sigma}_z\rangle &= \operatorname{Im}\,\chi(\beta) \quad (\text{for }\theta=\pi/2).
\end{align}
\vspace{-5ex}

Crucially, the measurement protocol for $\chi(\beta)$ requires strict phase synchronization between the bichromatic laser pulse that generates the state-dependent displacement and the RF pulses used to engineer the nonreciprocal coupling during motional-state preparation. An uncontrolled phase offset misaligns the phase-space axes between the preparation and measurement stages, rendering the entire measurement scheme invalid. We perform a calibration routine as follows. First, a coherent state is prepared by resonantly driving the motional mode with an RF pulse. Using a bichromatic laser pulse that produces the same displacement amplitude, we adjust its optical phase until the prepared coherent state is returned to the phase-space origin. This condition determines the relative phase between the laser field and the RF drive. Based on this reference, synchronizing the subsequent measurements becomes straightforward.

\subsection{Discrete Grid Sampling of the Characteristic Function}

The experimental sampling of the characteristic function is performed on a discrete polar grid,
\begin{equation}
	\beta = |\beta|\,\mathrm{e}^{\mathrm{i}\theta_{\beta}}.
\end{equation}
The radial displacement amplitude $|\beta|$ is controlled by varying the pulse duration of the state-dependent displacement operation, while the angular coordinate $\theta_{\beta}$ is set by the relative phase of the bichromatic field. At each grid point, the measurement protocol described in Sec.~\ref{Protocol} is executed, yielding either $\operatorname{Re}\chi(\beta)$ or $\operatorname{Im}\chi(\beta)$ from the measured $\langle\hat{\sigma}_z\rangle$, depending on the initial spin orientation. To mitigate the effects of slow frequency drifts, all repeated measurements at a given grid point are taken immediately after the corresponding calibration of the displacement amplitude and phase.

\begin{figure}[t]
	\centering
	\includegraphics[scale=0.35]{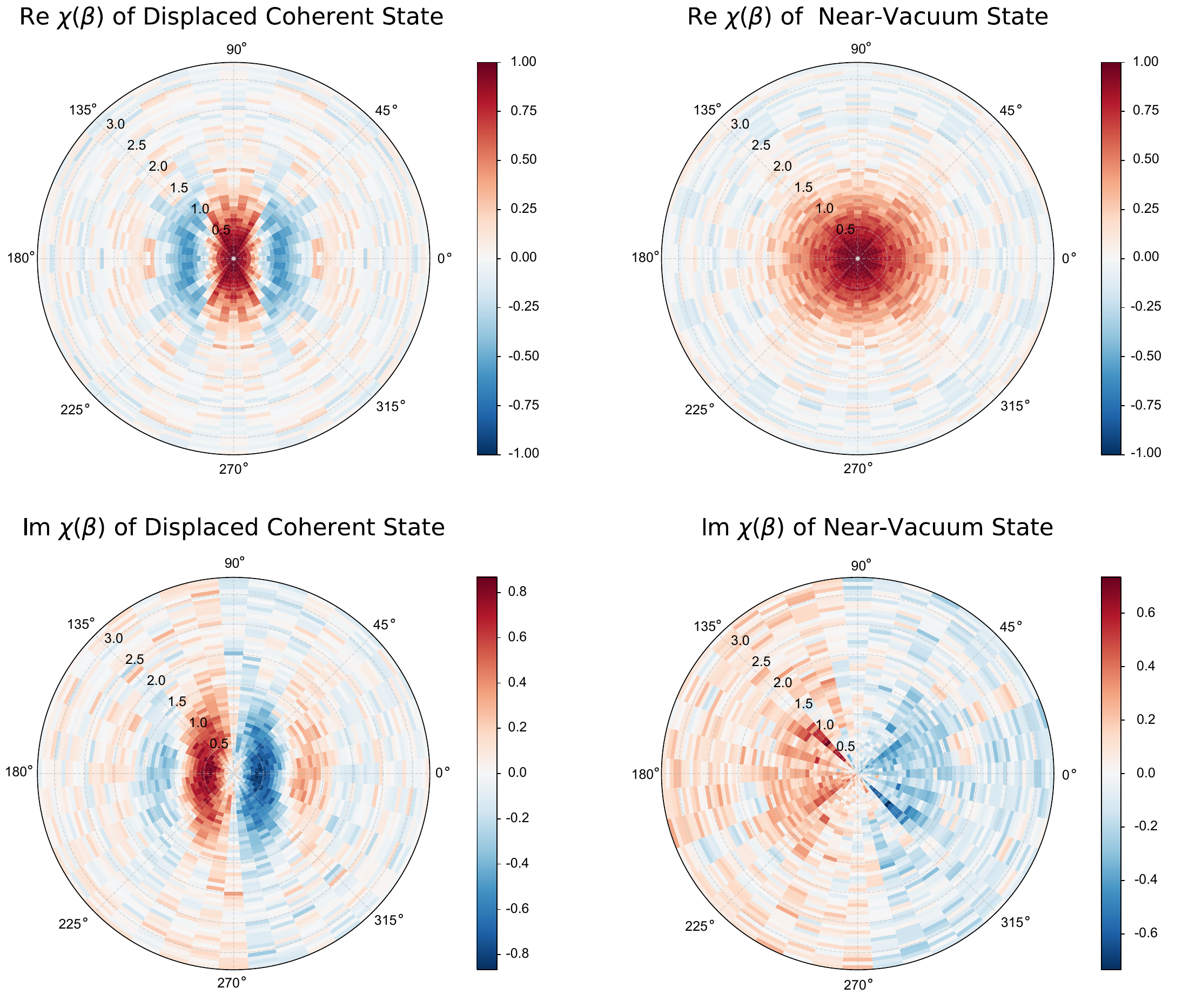}
	\caption{
		Experimental data of the characteristic function for the steady state of the nonreciprocal battery (a displaced coherent state) and for the state after discharge (a near-vacuum state) in Fig.~3.  
	}
	\label{FigS12}
\end{figure}

The Hermiticity of the motional density operator imposes the conjugate symmetry relation $\chi(-\beta) = \chi^{*}(\beta)$. Consequently, we experimentally measure $\operatorname{Re}\chi$ and $\operatorname{Im}\chi$ only over the right half of phase space; the values in the left half are obtained via
\begin{align}
	\operatorname{Re}\chi(|\beta|,\theta_{\beta}+\pi) &= \operatorname{Re}\chi(|\beta|,\theta_{\beta}), \\
	\operatorname{Im}\chi(|\beta|,\theta_{\beta}+\pi) &= -\operatorname{Im}\chi(|\beta|,\theta_{\beta}).
\end{align}
Figure~\ref{FigS12} shows the measured real and imaginary parts of the characteristic function for the steady state of the nonreciprocal battery (a displaced coherent state) and for the state after discharge (a near-vacuum state) in Fig.~3.

\subsection{Wigner Function Reconstruction via Fourier Transform}

Once the full characteristic function is obtained, the corresponding Wigner distribution $W(\alpha)$ is reconstructed via a two-dimensional Fourier transform,
\begin{equation}
	W(\alpha) = \frac{1}{\pi^2} \int \mathrm{d}^2\beta \; \chi(\beta) \exp(\alpha\beta^* - \alpha^*\beta),
\end{equation}
where $\alpha = x + \mathrm{i}p$, with $x$ and $p$ denoting the canonical position and momentum in phase space. To directly utilize the experimentally obtained characteristic function sampled on a discrete polar grid, this integral can be recast into the practical form
\begin{equation}
	W(x,p) = \frac{1}{\pi^2} \int_{0}^{2\pi} \mathrm{d}\theta_\beta \int_{0}^{\infty} \mathrm{d}|\beta| \; \chi(\beta) \, |\beta| \,  \exp\!\bigl[-2\mathrm{i}|\beta|(x \sin\theta_\beta - p \cos\theta_\beta)\bigr].
\end{equation}
Because the experimental sampling of $\chi(\beta)$ is necessarily truncated at a finite maximum radius $|\beta|_{\max}$, a suitable windowing function is applied during numerical integration to suppress high-frequency Gibbs ringing in the reconstructed Wigner function.

%
	
	\renewcommand{\theenumiv}{S\arabic{enumiv}}

\end{document}